\documentclass[epj]{svjour}
\usepackage{graphicx}
\usepackage{color}  
\usepackage[sort&compress,numbers]{natbib}
\usepackage{hyperref} 
\begin{document}
\title{Density dependence of the symmetry energy from neutron skin thickness in
finite nuclei}
%
\author{X. Vi\~nas\inst{1} \and M. Centelles\inst{1} \and X. Roca-Maza\inst{2}
\and M. Warda\inst{3}}
%
\institute{Departament d'Estructura i Constituents de la Mat\`eria and Institut
de Ci\`encies del Cosmos, Facultat de F\'{\i}sica, Universitat de Barcelona,
Diagonal 645, E-08028 Barcelona, Spain
\and
Dipartamento di Fisica, Universit\`a degli Studi di Milano and INFN, Via Celoria
16, I-20133 Milano, Italy
\and 
Katedra Fizyki Teoretycznej, Uniwersytet Marii Curie--Sk\l odowskiej ul.
Radziszewskiego 10, 20-031 Lublin, Poland}
%
\date{Received: date / Revised version: date}
%
\abstract{The density dependence of the symmetry energy around saturation
density, characterized by the slope parameter $L$, is studied using information
provided by the neutron skin thickness in finite nuclei. An estimate for $L$ is
obtained from experimental data on neutron skins extracted from antiprotonic
atoms. We also discuss the ability of parity-violating elastic electron
scattering to obtain information on the neutron skin thickness in $^{208}$Pb and
to constrain the density dependence of the nuclear symmetry energy.
The size and shape of the neutron density distribution of $^{208}$Pb predicted
by mean-field models is briefly addressed. We conclude with a comparative
overview of the $L$ values predicted by several existing determinations.
%
\PACS{{21.10.Gv, 21.65.Ef, 21.60.-n, 25.30.Bf, 26.60.-c}{Symmetry energy.
Neutron density. Neutron skin thickness. Parity-violating electron scattering.}
} 
} 
%
\maketitle

%
%
\section{Introduction}
\label{intro}
The nuclear symmetry energy is a fundamental quantity in nuclear physics and
astrophysics because it governs important properties of very small entities like
atomic nuclei ($R \sim 10^{-15}$ m) and of very large objects like neutron stars
($R \sim 10^{4}$ m) \cite{Ste05}. The symmetry energy represents the energy cost
per particle of changing all the protons into neutrons in symmetric nuclear
matter. In nuclear physics the symmetry energy has important consequences for
studies of the neutron distribution in stable and exotic nuclei
\cite{Bro00a,Bro00b,Bro00c,Hor01a,Hor01b,Die03,Ste05a,Gai12}, heavy-ion
collisions (HIC) \cite{Khoa96,Gai04,Bar05,Li08,She07,Fam06,Tsa09}, giant
and pygmy resonances \cite{Tri08,Kli07,Roca12,carbone10,Roca13}, etc. The
symmetry energy is also relevant for some aspects of fundamental physics because
it has implications for atomic parity non-conserving observables involved in
low-energy tests of the Standard Model and new physics \cite{Gue05,Sil05}. In
astrophysics the symmetry energy is very important for understanding different
properties of neutron stars, supernova explosions, and stellar nucleosynthesis
\cite{Ste05,Lat07,Klahn06,Khoa11}.

The Equation of State (EOS) in asymmetric nuclear matter can be expressed as
\begin{equation} 
e(\rho,\delta)= e(\rho,0) + S(\rho)\delta^2 + {\cal O}(\delta^4) ,
\label{eq1}
\end{equation}
where $\rho=\rho_n+\rho_p$ is the total nucleon density and
$\delta=(\rho_n-\rho_p)/\rho$ defines the relative asymmetry parameter. In
Eq.~(\ref{eq1}), $e(\rho,\delta)$ is the energy per particle in asymmetric
nuclear matter and $S(\rho)$ is the symmetry energy which is defined as
\begin{equation}
S(\rho) =\frac{1}{2} 
\left. \frac{\partial^2 e(\rho,\delta)}{\partial \delta^2} \right|_{\delta=0}.
\label{eq2}
\end{equation}
To a good accuracy $S(\rho)$ can be estimated as the difference between the EOS
of pure neutron matter and symmetric nuclear matter, i.e.,
\begin{equation}
S(\rho) \approx e(\rho,\delta=1) - e(\rho,\delta=0) .
\label{eq3} 
\end{equation}

It is both common and insightful to expand the EOS of asymmetric nuclear matter
around the saturation density $\rho_0$ using a few meaningful bulk parameters,
as in
 \begin{equation}
e(\rho,\delta=0)\simeq a_v
+ \frac{1}{2} K_v\epsilon^2 
\label{eq4a}
\end{equation}
and
\begin{equation}
 S(\rho)\simeq J - L\epsilon + \frac{1}{2} K_{\rm sym} \epsilon^2 ,
\label{eq4b}
\end{equation}
where the density shift is defined as $\epsilon=(\rho_0-\rho)/(3\rho_0)$. The
energy per particle $a_v$ and incompressibility $K_v$ in symmetric nuclear
matter at saturation density $\rho_0 \approx0.16$ fm$^{-3}$ are well constrained
by the measured data on densities, masses and collective excitations of nuclei.
They are around $a_v\!\approx\!-16$ MeV and $K_v\approx230\pm30$ MeV
\cite{Sto12}. The symmetry energy coefficient is defined as $J=S(\rho_0)$.
Experimental informations from nuclear masses, heavy-ion reactions, giant
resonances, and observational properties of neutron stars, constrain $J$ in a
range between 30 and 35 MeV \cite{Tsa12,Lat13}. These values are, in general,
predicted also by successful nuclear mean-field models.

\begin{figure*}
\label{fig1}
\begin{center}
\resizebox{0.75\textwidth}{!}
{\includegraphics{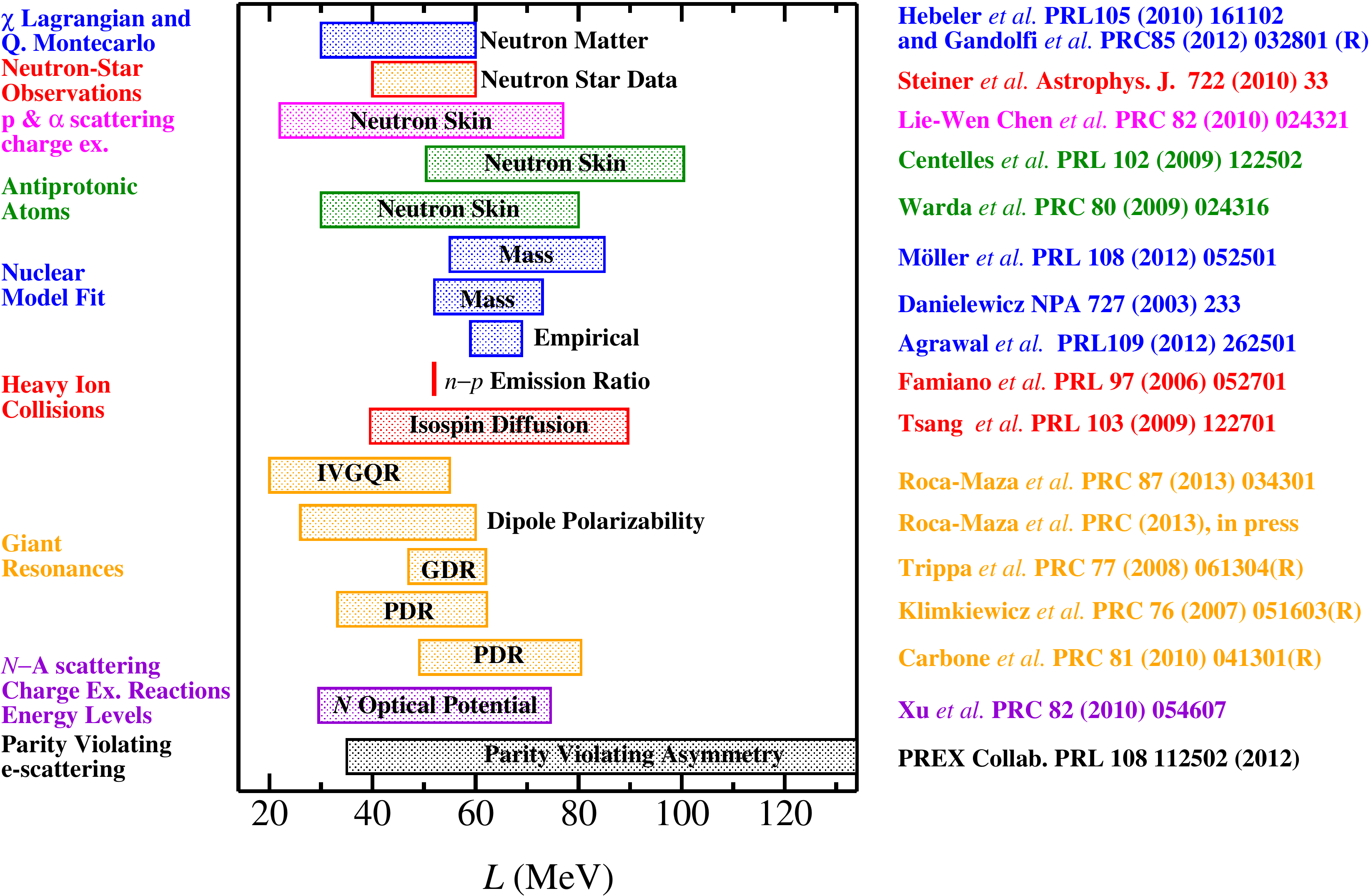}}
\end{center}
\caption{(Color online) Different available estimates of the slope parameter of the
symmetry energy $L$.}
\end{figure*}

However, the density dependence of the symmetry energy, characterized by the parameters
\begin{equation}
L=  \left. 3\rho\frac{\partial S(\rho)}{ \partial\rho}
\right|_{\rho_0} \label{eq5a}
\end{equation}
and
\begin{equation}
K_{\rm sym}= \left.
9\rho^2\frac{\partial^2 S(\rho)}{ \partial\rho^2} \right |_{\rho_0},
\label{eq5b}
\end{equation}
is much less certain and the predicted values differ largely among different
nuclear models (see e.g.\ Ref.\ \cite{Li08}). This is a consequence of the fact
that data of stable nuclei, used in the fits of the parameters of the models, do
not constrain in a very precise way the isovector sector of the effective
nuclear interaction. The $L$ value predicted by successful mean-field models
covers a wide range roughly between 20 MeV and 120 MeV. To extract this quantity
from available experimental data is not an easy task and usually some model
assumptions have to be made in order to connect the data with the parameter $L$.

A sample of different experimental results and theoretical calculations
used to estimate $L$ is displayed in Fig.~\ref{fig1}. We will overview them in
some more detail later in Section \ref{sec:2.4}. For in-depth accounts of
several of the methods and determinations shown in Fig.~\ref{fig1}, see the
related contributions in the present volume (also see Refs.~\cite{Tsa12,Lat13}).
Fig.~\ref{fig1} illustrates that although very
different observables and methods have been used to extract $L$ and different
values have been obtained, there is a common range, approximately between 45 and
70 MeV, where the different estimates of $L$ have a large overlap. Such a range
can be considered as a reliable estimate of the $L$ parameter. This range of
values agrees also with the range reported in Ref.~\cite{Sto12}.

On the basis of our findings in recent years \cite{Cen09,War09,warda10,cen10,Roc11}, in the next sections we will discuss two of these estimates of the $L$ parameter. Namely, we will study the constraints that can be derived from the analysis of neutron skins in antiprotonic atoms, and the constraints that may be provided by parity-violating elastic electron scattering in a heavy neutron-rich nucleus such as $^{208}$Pb. We also discuss the size and shape of the neutron density distribution of $^{208}$Pb predicted by mean-field models and its dependence with the symmetry energy. A summary and outlook are given to conclude the paper.

\section{Insights into the neutron skin thickness from the nuclear droplet model}
\label{sec:1}
\subsection{Theory}
\label{sec:2}

\begin{figure}[t]
\includegraphics[width=1.0\columnwidth,clip=true]{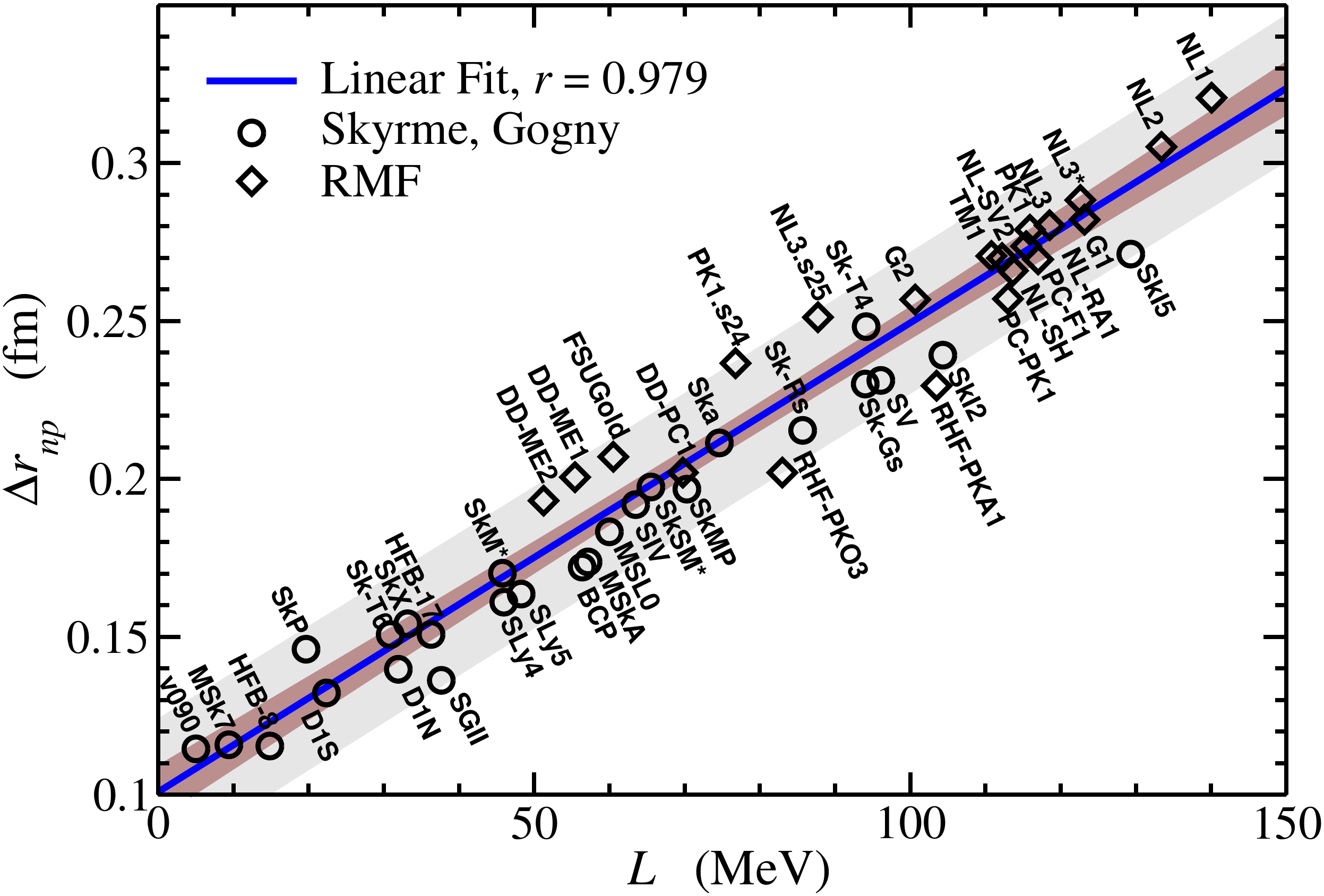}
\caption{\label{fig2} (Color online) Neutron skin thickness $\Delta r_{np}$ of
$^{208}$Pb against the slope parameter of the symmetry energy at saturation
density $L$. The predictions of diverse nuclear energy density functionals
(including Skyrme and Gogny forces and relativistic mean-field models) are
shown. The linear correlation coefficient of the results for $\Delta r_{np}$ and
$L$ is $r=0.979$. A linear fit gives $\Delta r_{np} = 0.101 + 0.00147 L$, with
$\Delta r_{np}$ in fm and $L$ in MeV. The thinner and thicker shadowed regions
represent, respectively, the 95\%-confidence band and the 95\%-prediction band
of the linear regression \cite{regres}. Figure adapted from Ref.~\cite{Roc11}.}
\end{figure}

The neutron skin thickness (NST) of a nucleus is conventionally defined as the difference between the neutron and proton rms radii of the nucleus:
\begin{equation}  
\Delta r_{np}= \langle r^2 \rangle_n^{1/2} - \langle r^2 \rangle_p^{1/2}. 
\label{eq5} 
\end{equation} 
It is known from almost fifteen years ago that in a heavy nucleus such as
$^{208}$Pb the NST calculated with successful mean-field models that reproduce
the binding energy and charge radius of this nucleus is linearly correlated with
the slope of the symmetry energy at saturation computed with the same model
\cite{Bro00a,Bro00b,Bro00c}. This correlation can be clearly seen in
Fig.~\ref{fig2}, where Hartree-Fock or Hartree calculations of the NST of
$^{208}$Pb with different Skyrme, Gogny and relativistic mean-field models is
displayed as a function of the parameter $L$ defined in Eq.\ (\ref{eq5a}), which
is proportional to the slope of the symmetry energy at saturation. The
linear fit of the results of Fig.~\ref{fig2} gives
\begin{equation}
\Delta r_{np} \;{\rm (fm)} \;=\; 0.101 + 0.00147 L \;{\rm (MeV)} \;.
\label{fitrnpL}
\end{equation}
Information about the models of the figure is provided in Section
\ref{resultsI}.

A possible explanation of the correlation between $\Delta r_{np}$ and $L$ on the
basis of the nuclear Droplet Model (DM) \cite{MSa,MSb,MSc} has been reported in
Ref.~\cite{Cen09}. In the DM, the symmetry energy coefficient of a nucleus with
mass number $A$ is obtained as \cite{MSa,MSb,MSc}
\begin{equation}
\label{eq6}
a_{\rm sym}^{\rm DM}(A)= \frac{J}{1+x_A},
\quad \mbox{with }\,x_A= \frac{9J}{4Q}A^{-1/3},
\end{equation}
where $J$ is the bulk symmetry energy coefficient defined previously. The
corresponding symmetry energy contribution to a finite nucleus is given by
\cite{MSa,MSb,MSc}
\begin{equation}
E_{\rm sym}^{\rm DM}(A) = a_{\rm sym}^{\rm DM}(A) \,(I+x_A I_C)^2 A \,,
\label{eq7}
\end{equation}
where $I=(N-Z)/A$ is the overall relative neutron excess. The quantity
$I_{\rm C}= e^2 Z/(20J R)$, where $R=r_0 A^{1/3}$ is the nuclear radius, is a
correction due to the Coulomb interaction \cite{MSa,MSb,MSc}.

The coefficient $Q$ of Eq.\ (\ref{eq6}) is called the surface stiffness
coefficient. It measures the resistance of neutrons against separation from
protons to form the neutron skin. In the context of the DM and under the
assumption that a finite nucleus is a leptodermous finite Fermi system, having a
wide bulk part and a narrow surface region where the density rapidly drops to
zero, nuclear surface properties such as e.g.\ the coefficient $Q$ can be
extracted from semi-infinite nuclear matter calculations
\cite{MSa,MSb,MSc,Far81,Bra85,Kol85,Tre86}. The semi-infinite geometry
is free of finite-size effects, Coulomb, and shell effects. Quantal or
semiclassical extended Thomas-Fermi (ETF) calculations of asymmetric
semi-infinite nuclear matter (ASINM) can be performed  to compute $Q$
\cite{Far81,Bra85,Kol85,Tre86,Cen98,Estal99}. Here we perform ETF
calculations of ASINM, since the semiclassical ETF method does not present
Friedel oscillations found in the densities of a Hartree-Fock or Hartree
calculation of ASINM \cite{Cen93,Dan09}. A semiclassical approach does not have
quantal shell effects and hence is consistent with the DM concept of an average
description of nuclear properties \cite{Bra85}. Details of our ETF calculations
are given in \cite{Cen98} and in the appendix of \cite{War09}. For a comparison
of some DM predictions against self-consistent ETF results in finite nuclei, see
Section \ref {sec:2.3} below.

A link between finite nuclei observables, described by the DM, and certain
quantities such as $J$ and $L$ that define the EOS in nuclear matter around
saturation, may be obtained from the following empirical consideration
\cite{Cen09,War09}. The DM symmetry energy coefficient $a_{\rm sym}^{\rm DM}(A)$
of a heavy nucleus like $^{208}$Pb equals the value of the symmetry energy in
infinite matter $S(\rho)$ at a subsaturation density near 0.1~fm$^{-3}$ when
computed in mean-field models. More recently, this relation has been discussed
by Liu et al.\ from measured nuclear masses \cite{Liu10}, by Chen et al.\ in the
Skyrme-Hartree-Fock theory \cite{Lie11,Zhang2013}, and by Dong et al.\ in a
variety of mean-field interactions \cite{Dong2012,Dong2013}. For several nuclear
models, we display in Table 1 the bulk symmetry energy coefficient $J$, the DM
symmetry energy coefficient (\ref{eq6}), and the value of the density $\rho$
that fulfils $S(\rho)=a_{\rm sym}^{\rm DM}(A)$ for $A=208$. Of course, though it
is around 0.1 fm$^{-3}$, there is a certain variation in the value of $\rho$
where the condition is fullfilled exactly.

\begin{table}
\label{Table1}
\caption{Coefficients $J$ and $a_{\rm sym}^{\rm DM}(A)$ and density $\rho$ that
fulfils $S(\rho)=a_{\rm sym}^{\rm DM}(A)$ for $A=208$ in several nuclear models.
The values of $a_{\rm sym}^{\rm DM}(A)$ and $S(\rho)$ have been computed
with Eqs.\ (\ref{eq6}) and (\ref{eq2}), respectively.}
\begin{center}
\begin{tabular}{lccc}
\hline\noalign{\smallskip}
Model & $J$ (MeV) &
$a_{\rm sym}^{\rm DM}$ (MeV) & $\rho$ (fm$^{-3}$) \\ \hline
NL3           &37.4&25.8&0.103\\
NL-SH         &36.1&26.0&0.105\\
FSUGold       &32.6&25.4&0.099\\
TF            &32.6&24.2&0.094\\
SLy4          &32.0&25.3&0.100\\
SkX           &31.1&25.7&0.103\\
SkM*          &30.0&23.2&0.101\\
SIII          &28.2&24.1&0.093\\
SGII          &26.8&21.6&0.104\\
\hline\noalign{\smallskip}
\end{tabular}
\end{center}
\end{table}

Within the DM the NST of a nucleus is given by \cite{Mye80}
\begin{equation}
\Delta r_{np}^{\rm DM} = \sqrt{\frac{3}{5}} \left[t - \frac{e^2Z}{70J} +
\frac{5}{2r_0 A^{1/3}}(b^2_n - b^2_p)\right].
\label{eq9}
\end{equation}
The quantity $t$ is the distance between the neutron and proton mean surface
locations. The correction $e^2Z/70J$ is due to the Coulomb interaction, and
$b_n$ and $b_p$ are the surface widths of the neutron and proton density
profiles. Eq.~(\ref{eq9}) shows that the NST in a heavy nucleus may be formed by
two basic mechanisms. One of them is due to the separation between the neutron
and proton mean surfaces, which is a bulk effect. The other way to form the NST
is a surface effect due to the fact that the widths of the neutron and proton
surfaces may be different.

In the DM the bulk contribution $t$ to the NST can be written as  
\begin{eqnarray}
t &=& \frac{3}{2} r_0 \frac{J}{Q} \frac{1}{1+x_A}(I -I_C)
\nonumber \\
&=& \frac{2r_0}{3J}\big[J -a_{\rm sym}^{\rm DM}(A)\big]A^{1/3}(I - I_C),
\label{eq10}
\end{eqnarray}
where use is made of Eq.(\ref{eq6}) to establish the second equality. From the
first equality in (\ref{eq10}), it is seen that the leading contribution to $t$
in a large nucleus is the term $\frac{3}{2} r_0 (J/Q) I$. Thus, the DM suggests
a correlation between $\Delta r_{np}$ and $J/Q$ in heavy nuclei. The second
equality in (\ref{eq10}) shows that the DM establishes a direct relation between
the surface symmetry energy coefficient $\big[J -a_{\rm sym}^{\rm
DM}(A)\big]A^{1/3}$ and the bulk part of the NST $t$. The standard version of
the DM assumes $b_n=b_p=1$ fm \cite{MSa,MSb,MSc}. Taking into account the
relation between the symmetry energy in infinite matter and the symmetry energy
coefficient of a finite nucleus in the DM model ($S(\rho) \approx a_{\rm
sym}^{\rm DM}(A)$ at a subsaturation density $\rho$), we can write
 \begin{equation}
t= \frac{2r_0}{3J}L \left(1 -\epsilon
\frac{K_{\rm sym}}{2L}\right)\epsilon A^{1/3} (I-I_C). \label{eq11}
\end{equation}
The result explicitly shows the imprint of the density dependence of the
symmetry energy on the NST. The leading proportionality in Eq.(\ref{eq11})
between the bulk part of the NST $t$ and the slope parameter of the symmetry
energy $L$, explains the observed correlation of $\Delta r_{np}$ with $L$ in a
heavy nucleus such as $^{208}$Pb \cite{Bro00a,Bro00b,Bro00c,Cen09} displayed in
Fig.~\ref{fig2}. Regarding Eq.\ (\ref{eq11}) it should be pointed out that the
corrective term with $K_{\rm sym}$ has a limited influence on the result as far
as $\epsilon$ is small.

\subsection{Constraints from empirical data on the neutron skin thickness from
antiprotonic atoms}
\label{sec:2.2}

Slow-enough antiprotons can form with an atomic nucleus a hydrogen-like
atom. When the antiproton annihilates with a nucleon, pions are produced that
may miss the nucleus and a residue is left with a neutron or proton less. From
the analysis of the yields, information about the neutron distribution in the
nucleus can be derived (cf.\ \cite{Trz01,Jas04} and references therein). A
second method measures antiprotonic X-rays, which allow to find the atomic
level widths and shifts due to the strong interaction \cite{Trz01,Jas04,Klo07}.
Combining the results from the two techniques, the difference $\Delta r_{np}$
between the neutron and proton rms radii can be estimated if the charge density
of the nucleus is known~\cite{Trz01,Jas04,Klo07}.
The NST of 26 stable nuclei, ranging from $^{40}$Ca to $^{238}$U, was extracted
from experiments with antiprotons at CERN by Trzci{\'n}ska et al.\
\cite{Trz01,Jas04}. The data roughly follow a linear trend with the relative
neutron excess $I$ of the nucleus that can be fitted by \cite{Trz01,Jas04}
\begin{equation}
\Delta r_{np}= (0.90 \pm 0.15) I + (-0.03 \pm 0.02)~{\rm fm} .
\label{apfit}
\end{equation}
It is to be mentioned that the NST values derived from antiprotonic atoms
have relatively large uncertainties, as it can be seen in Fig.~\ref{fig2}, and
sometimes the results deduced from different experiments are not completely
concordant with each other \cite{Kra04,Jas04}. However, the NST data of
Refs.~\cite{Trz01,Jas04} are the largest available set of $\Delta
r_{np}$ values uniformly measured with the same experimental technique all
over the periodic table ($40 \leq A \leq 238$). Moreover, all of the nuclei are
stable isotopes.

\begin{figure} 
\resizebox{0.5\textwidth}{!} 
{\includegraphics[clip=true]{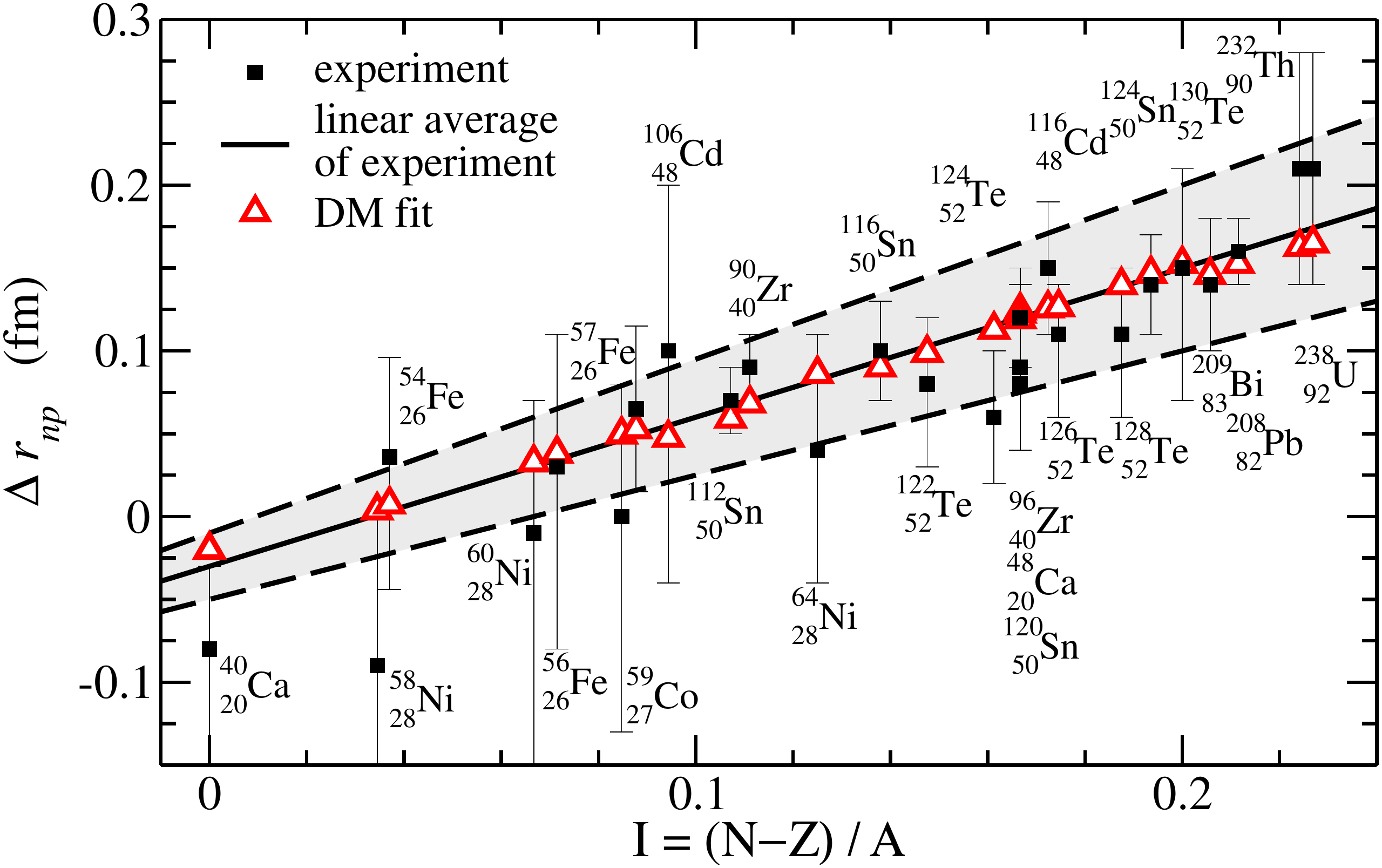}}
\caption{\label{fig3} (Color online) Neutron skin thickness as a function of
the asymmetry parameter of the nucleus $I=(N-Z)/A$ for the nuclei measured in
the experiments with antiprotonic atoms. The droplet model results have been
obtained with Eqs.~(\ref{eq9}) and (\ref{eq10}) (see text for details). Figure
adapted from Ref.~\cite{Cen09}.}
\end{figure}

We next explore~\cite{Cen09} the constraints on the density dependence of the
symmetry energy by fitting Eq.~(\ref{eq9}) to the neutron skins measured in
Refs.\ \cite{Trz01,Jas04}. To have a large set of skins across the mass table 
at our avail is very helpful because the result of the fit is driven mainly by
the average trend and is less contingent on the data uncertainties. In our
fit the explicit dependence of the NST on $L$ is provided by Eq.\ (\ref{eq11}).
In this analysis we parametrize the symmetry energy as
\begin{equation}
S(\rho) = J \, \big(\rho/\rho_0\big)^\gamma ,
\label{eq11a}
\end{equation}
with $J$ the symmetry energy coefficient. This expression has been often used in
studies of HIC assuming $J=31.6$ MeV
\cite{Li05a,Li05b,Li08,She07,Fam06}. To start with, we set $b_n=b_p$ as has been
done in the standard DM \cite{MSa,MSb,MSc,Mye80,Swi05} and in the analysis of
data in Ref.~\cite{Dan03}. In this way we find $L=75\pm25$ MeV
($\gamma=0.79\pm0.25$) from the fit. The quoted uncertainty is from the window
of the linear averages of experiment. The $L$ value and its uncertainty obtained
from the DM NST formula with a vanishing surface width correction is compatible
with the $L$ values predicted from isospin diffusion and isoscaling observables
in HIC in Refs.~\cite{Li05a,Li05b,Li08,She07}. In Fig.~\ref{fig3} we display by
open triangles the NST predicted by our DM calculation for the 26 nuclei; the
results nicely reproduce the average trend of the experimental data.

\subsection{Contribution of the nuclear surface to the neutron skin thickness}
\label{sec:2.3}

The results just discussed have been obtained assuming same neutron and proton
surface widths (standard version of the DM). However, the analysis of
mean-field nucleon densities carried out in Ref.\ \cite{Miz00} by means of the
Helm model, pointed out that the neutron and proton density profiles differ not
only in the mean location of their surfaces but also in their surface widths.

\begin{figure}
\resizebox{0.45\textwidth}{!}
{\includegraphics[clip=true]{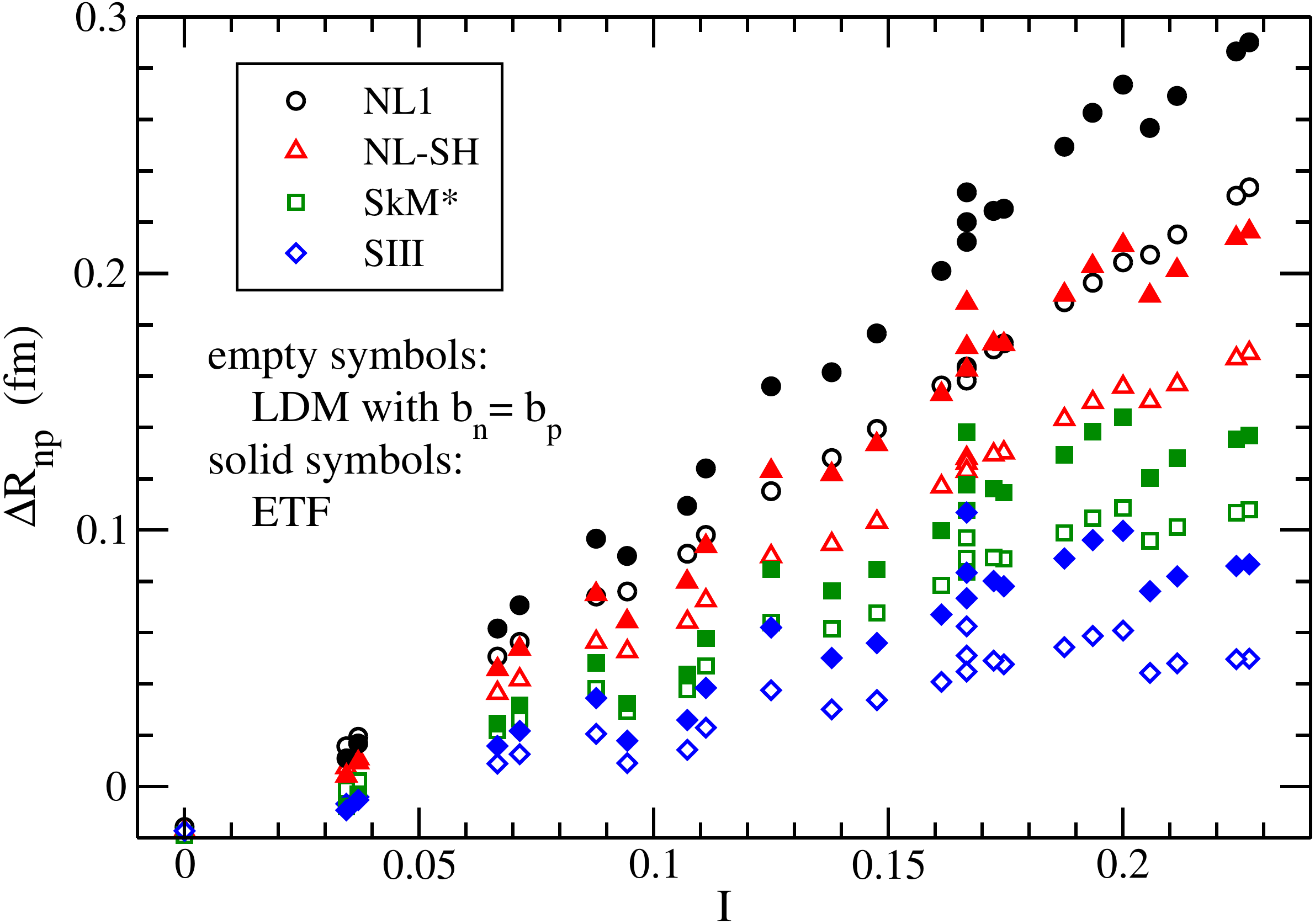}}
\caption{(Color online) The neutron skin thickness predicted by the standard
version of the DM (Eq.\ (\ref{eq9}) with $b_n=b_p$) is compared with the result
from self-consistent ETF calculations of finite nuclei, in four mean-field
models. The nuclei considered are those investigated in antiprotonic atoms in
Refs.~\cite{Trz01,Jas04} and have masses $40 \leq A \leq 238$. Figure taken from
Ref.\ \cite{War09}.
\label{fig4}}
\end{figure}

A similar situation occurs when the NST is calculated in the DM including the
surface width correction in Eq.~(\ref{eq9}). First, in Fig.~\ref{fig4} we
display by empty symbols the NST predicted by the DM with $b_n=b_p$, for the
same nuclei measured in antiprotonic atoms, computed with Eqs.\ (\ref{eq9}) and
(\ref{eq10}) using the SIII and SkM* Skyrme forces and the NL-SH and NL3
relativistic mean field (RMF) parameter sets. These values are compared in
Fig.~\ref{fig4} with the values
obtained from {\sl self-consistent} ETF calculations in {\sl finite} nuclei
\cite{Cen90,Cen93} (filled symbols). The reason for comparing DM with ETF
instead of with Hartree-Fock or Hartree results, lies on the fact that both, 
DM and ETF, are free of shell effects that could mask the comparison. We
have calculated the ETF values of the NST by using Eq.\ (\ref{eq5}) with the rms
radii of the self-consistent ETF neutron and proton densities of each isotope.

Two conclusions can be underlined from Fig.~\ref{fig4}. First, the predictions
of the DM with $b_n=b_p$ systematically undershoot the ETF values of the NST
computed in finite nuclei. Second, for a given nucleus the difference between
the ETF value of $\Delta r_{np}$ and the value predicted by the DM with
$b_n=b_p$ is slightly larger in the RMF parameter sets than in the Skyrme
forces. These facts denote that in mean-field models the surface width
correction to the DM formula for the NST does not vanish, and that it has some
dependence on the value of the $J/Q$ ratio.

\begin{figure}
\resizebox{0.45\textwidth}{!}
{\includegraphics[clip=true]{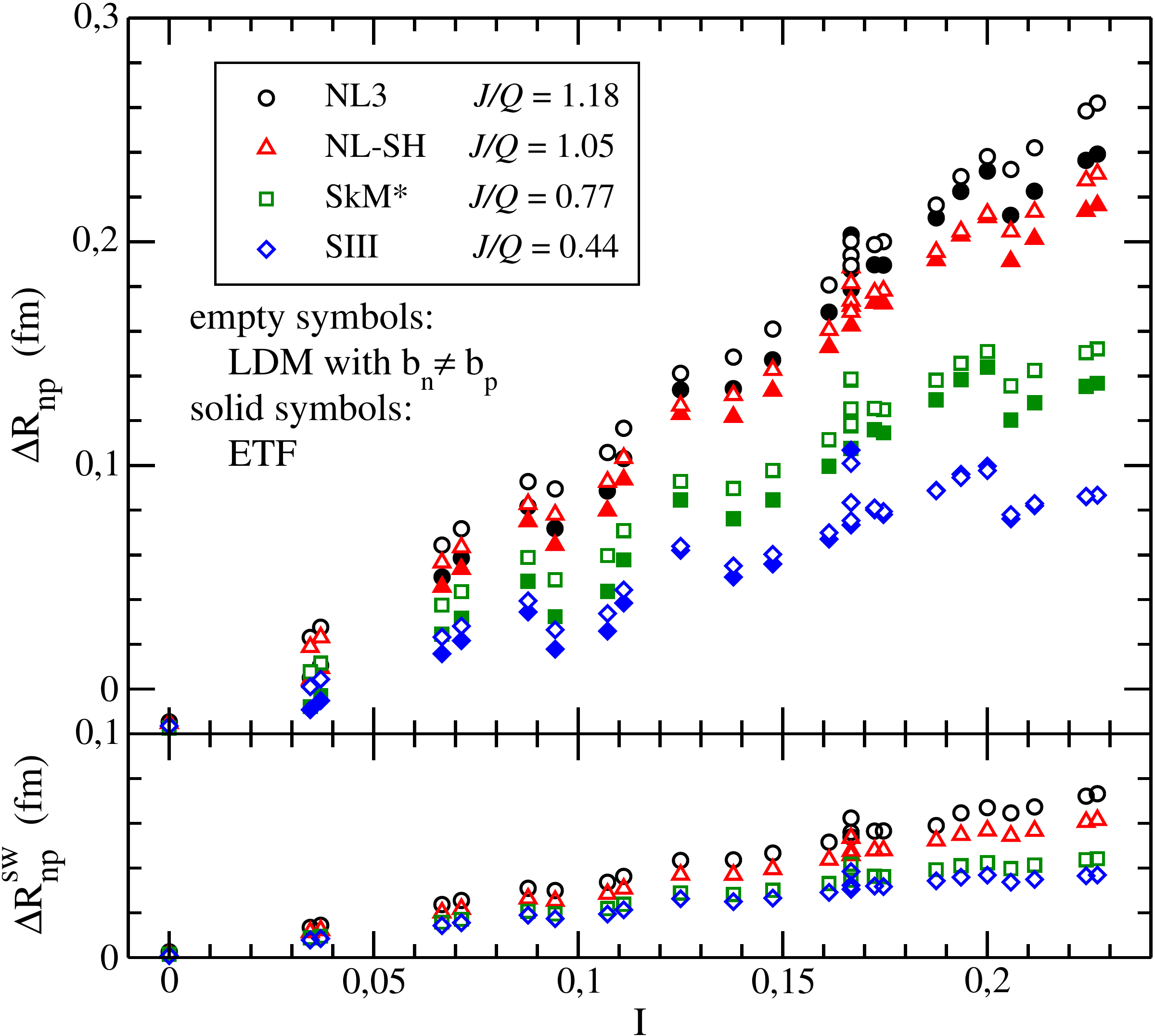}}
\caption{(Color online) Upper panel: the same as in Fig.~(\ref{fig4}) but here
the DM values include the surface width correction $\Delta r_{np}^{sw}$
[Eq.~(\ref{eq13})] with $b_n$ and $b_p$ obtained from ASINM as described in the
text. Lower panel: surface width correction $\Delta r_{np}^{sw}$ (the vertical
scale proportionality is the same as in the upper panel). Figure taken from
Ref.\ \cite{War09}.
\label{fig5}}
\end{figure}

In order to compute Eq.\ (\ref{eq9}) including the surface width correction, one
needs to estimate the values of the $b_n$ and $b_p$ surface widths; we proceed
as before and compute $b_n$ and $b_p$ by performing ETF calculations of ASINM
\cite{Bra85,Cen98,Cen09,War09} (see appendix of \cite{War09} for further
details). Once $b_n$ and $b_p$ are known, their
contribution to the NST is given by (see Eq.(\ref{eq9}))
\begin{equation}
\Delta r_{np}^{sw}=\sqrt{\frac{3}{5}}
\frac{5}{2R}\left(b_n^2-b_p^2\right)\,.
\label{eq13}
\end{equation}
The values of $\Delta r_{np}^{sw}$ for the nuclei considered in Fig.~\ref{fig4}
are displayed in the bottom panel of Fig.~\ref{fig5}. We see that $\Delta
r_{np}^{sw}$ shows for each nuclear interaction an increasing linear trend with
$I=(N-Z)/A$. We display the DM NST including the $\Delta r_{np}^{sw}$ correction
in the top panel of Fig.~\ref{fig5} by empty symbols. Note that these results
correspond to adding $\Delta r_{np}^{sw}$, shown in the bottom panel of this
figure, to the DM values we have displayed in Fig.~\ref{fig4}.  We observe in
the upper panel of Fig.~\ref{fig5} a quite good agreement between the new DM
predictions and the {\sl self-consistent} ETF values in {\sl finite}
nuclei computed with the same interaction, after including the $\Delta
r_{np}^{sw}$ correction in the DM formula.

The lower panel of Fig.~\ref{fig5} suggests that $\Delta r_{np}^{sw}$ has, for
each nucleus, a certain increasing trend with the $J/Q$ ratio. As discussed in
more detail in Ref.~\cite{War09}, it is possible to fit $\Delta r_{np}^{sw}$ by
means of a law $\sigma^{sw} I$, which defines the slope $\sigma^{sw}$ of $\Delta
r_{np}^{sw}$ with respect to $I$. This slope, as a function of the $J/Q$ ratio
for different interactions, lies inside a band limited by two straight lines,
each line corresponding to the equations $\sigma^{sw} = 0.3 J/Q +0.07$ fm and
$\sigma^{sw} = 0.3 J/Q - 0.05$ fm (see Fig.~4 of Ref.~\cite{War09}).

In summary, we have shown that to leading order both the bulk part $t$
(\ref{eq10}) and the surface width term $\Delta r_{np}^{sw}$ (\ref{eq13}) of the
DM NST are basically driven by the value of the $J/Q$ ratio. This fact suggests
to fit the experimental $\Delta r_{np}^{exp}$ data by means of the following DM
inspired ansatz:
\begin{equation}
\label{eq14}
\Delta r_{np} =
\sqrt{\frac{3}{5}}\left( \,t - \frac{e^2 Z}{70 J} \right)
+ \bigg( 0.3 \frac JQ + c \bigg)I ,
\end{equation}
where $t$ is given by Eq.\ (\ref{eq10}). The second term is the surface width
correction parametrized as explained before. With the ansatz (\ref{eq14}) we
have constrained the ratio $J/Q$ by a fit to the skin data in antiprotonic atoms
\cite{Trz01,Jas04}. Both Eq.\ (\ref{eq14}) and $t$ of Eq.\ (\ref{eq10}) depend
on the $J$ coefficient and the nuclear matter radius $r_0$. We fix
them to the empirical values $J=31.6$ MeV and $r_0=1.143$ fm (the latter
corresponds to a saturation density of 0.16 fm$^{-3}$). We consider $c=0.07$ fm
and $c=-0.05$ fm in Eq.\ (\ref{eq14}) to simulate the upper and lower bounds of
the window of the theoretical predictions for $\sigma^{sw}$ in the nuclear
mean-field models. More details of the fit can be found in Ref.~\cite{War09}.

The fits to experiment give $J/Q = 0.667 \pm 0.047$ with $c=0.07$ fm and $J/Q =
0.791 \pm 0.049$ with $c=-0.05$ fm (i.e., a total range $0.62 \le J/Q \le
0.84$). The quoted uncertainties in $J/Q$ correspond to one standard deviation
associated with the fit. Both extractions of $J/Q$, for $c=0.07$ and $c=-0.05$
fm, predict basically the same total values of the NST, which lie close
\cite{War09} to the average trend $\Delta r_{np}= (0.90 \pm 0.15) I + (-0.03 \pm
0.02)$ fm \cite{Trz01,Jas04} of the experimental data. However, if one looks
at the splitting of the NST into the bulk part $t$ and the surface width part
$\Delta r_{np}^{sw}$, it is different in the two cases.

\begin{figure}
\resizebox{0.5\textwidth}{!}
{\includegraphics[clip=true]{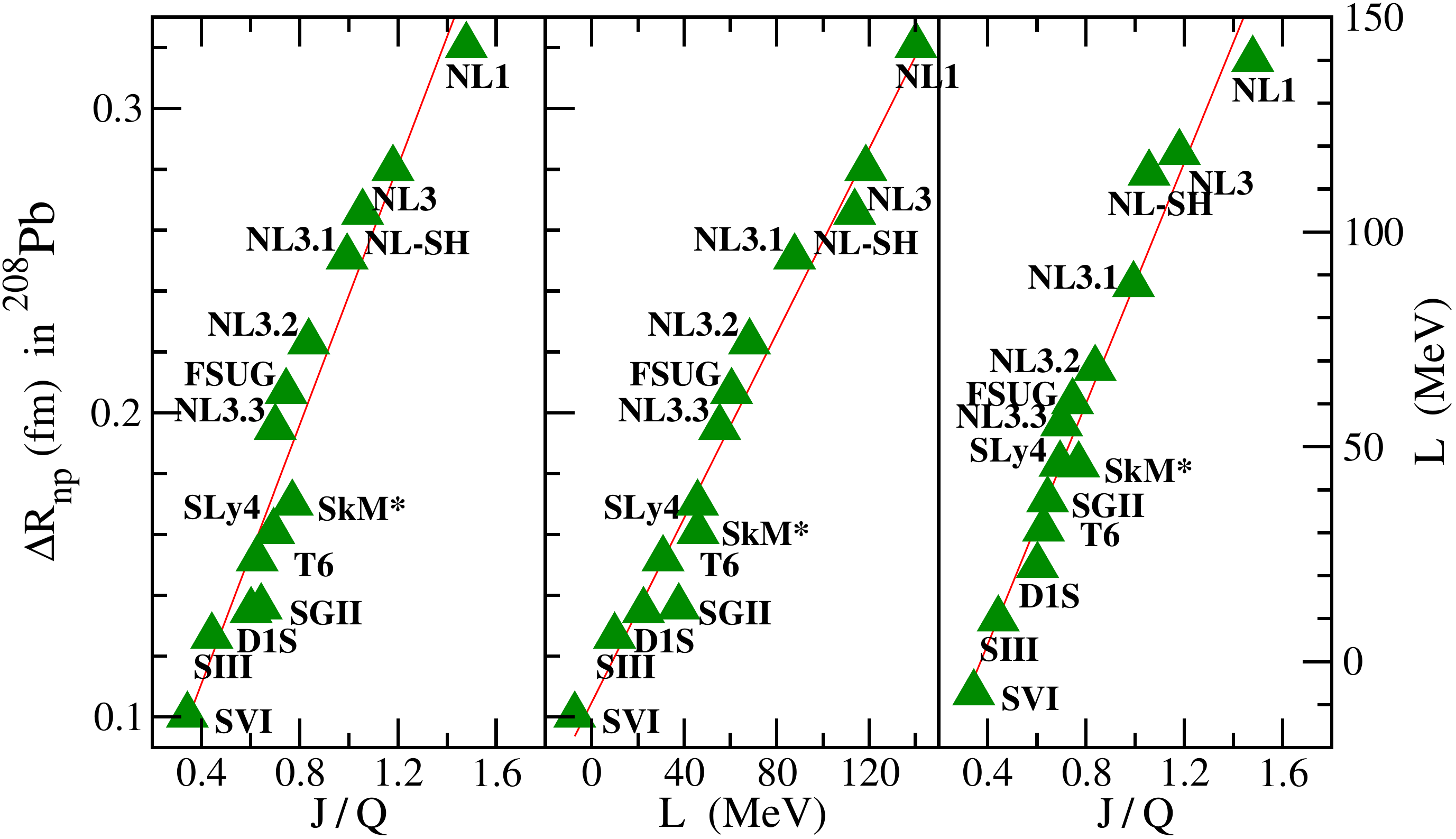}}
\caption{\label{fig6} (Color online) Correlation between the neutron skin
thickness $\Delta r_{np}$ in $^{208}$Pb and the ratio $J/Q$ (left panel) and
between $\Delta r_{np}$ in $^{208}$Pb and the slope of the symmetry energy at
saturation $L$ (middle panel) for several mean-field models.  The correlation
between $L$ and the ratio $J/Q$ is also shown (right panel).  The $\Delta
r_{np}$ values displayed in this figure have been computed with Eq.~(\ref{eq5})
from the rms radii of quantal self-consistent calculations with the mean-field
models. Figure taken from Ref.\ \cite{War09}.}
\end{figure}

As mentioned, the leading contribution to $t$ in large nuclei is the term
$\frac{3}{2} r_0 (J/Q) I$. Thus, the DM suggests that one can also expect a
correlation between $\Delta r_{np}$ and $J/Q$ in heavy nuclei. We illustrate
this fact in the left panel of Fig.~\ref{fig6}. The NST of $^{208}$Pb depicted
in this figure has been obtained from {\sl self-consistent quantal} calculations
with Skyrme, Gogny, and RMF models \cite{War09}. In the same figure the
correlation between the NST in $^{208}$Pb and the slope of the symmetry energy
at saturation $L$, discussed previously, is displayed to illustrate the expected
$L$-vs-$J/Q$ linear correlation shown in the rightmost panel of Fig.~\ref{fig6}.
We have checked that a linear fit of $L$ against $J/Q$ depends rather weakly on
the models selected to do the fit. We find that the fit lies in a range between
$L=139 J/Q - 52$ MeV and $L= 150 J/Q - 57$ MeV\@. Considering this result and
the constraint $0.6 \le J/Q \le 0.9$ discussed above, we find that $L$ is
comprised between 31 MeV and 78 MeV\@. Therefore, our estimate (which takes into
account the surface width correction in $\Delta r_{np}$) for the $L$ parameter
basically lies in a range $30 \le L \le 80$ MeV.

\section{Parity-violating electron scattering and neutron skin thickness}
\label{secpves}

In the previous section we have studied the constraints on the nuclear symmetry
energy that may emerge from consistently measuring the neutron skin thickness in
a wide range of nuclei across the mass table with a hadronic probe. The results
are very encouraging if improved accuracy could be achieved and, in our opinion,
they are supportive of pursuing further systematic experimental determinations
of neutron skins with hadronic probes.

In the following, we will address the complementary scenario where one explores
the constraints on the nuclear symmetry energy that may be obtained from a
high-precision measurement of the neutron skin thickness performed in only one
single nucleus (${}^{208}$Pb in our case) by using a leptonic probe.

Indeed, from Fig.~\ref{fig2} we see immediately that the predictions of the
nuclear mean-field models for the NST of the heavy
neutron-rich nucleus ${}^{208}$Pb vary by almost a factor of 3, and that the
slope parameter of the symmetry energy $L$ spans more than
an order of magnitude between the extreme cases. In spite of such
discrepancies, the mean field predictions for $\Delta r_{np}$ and $L$ show a
strong correlation (which in the previous section we have been able to
understand in terms of macroscopic arguments). Thus, an accurate measurement of
the NST (or of the neutron radius) of ${}^{208}$Pb, in addition to being of
fundamental importance by itself, would provide a unique constraint on the
density dependence of the nuclear symmetry energy at saturation.
Hence, it may be highly valuable to pursue new experiments to obtain
information about the neutron distribution in a heavy nucleus as much as
possible free from the uncertainties inherent in the strong interaction.

The Lead Radius Experiment (PREX) collaboration \cite{prex1,prex2,prexprl12} at
the Jefferson Lab has been working with the aforementioned purpose. PREX aims
to determine the neutron radius in $^{208}$Pb to 1\% accuracy by measuring the
parity-violating asymmetry ($A_{\rm pv}$) at low momentum
transfer in polarized elastic electron scattering performed at a single angle.
To reach a 1\% accuracy in the neutron radius, the accuracy needed in $A_{\rm
pv}$ has been estimated to be of a 3\%. The first run of
PREX was successfully completed in 2010 and demonstrated the validity of this
electroweak technique. The accuracy in $A_{\rm pv}$, though, was about 9\%
instead of 3\% \cite{prexprl12} mainly because of limited statistics in the
measurement. To attain the 3\%-accuracy goal in $A_{\rm pv}$, a new run
PREX-II will be performed \cite{prexii}. Recently, the motivations and
optimal conditions for measuring the parity violating asymmetry in other nuclei
such as $^{48}$Ca and $^{120}$Sn have been studied in Ref.~\cite{ban12} and
Refs.~\cite{liu12,liu13}. Indeed, a new proposal for performing parity-violating
electron scattering (PVES) on $^{48}$Ca has been approved
at the Jefferson Lab (CREX experiment) \cite{crex13}. Moreover,
high-precision PVES measurements of neutron radii may become feasible in the
future at the new MESA facility in Mainz~\cite{sfie13}.

For detailed informations on the PREX and CREX experiments and their
physics, see the contribution to this volume by Horowitz, Kumar, and Michaels.
Here, we summarize some results for completeness.
The reported value for $A_{\rm pv}$ from the first run of PREX has been
\cite{prexprl12}
\begin{equation}
  A_{\rm pv}=0.656\pm(0.060)_{\rm stat} \pm (0.014)_{\rm syst}\;{\rm ppm} \;,
\label{apvdat}
\end{equation}
at an average momentum transfer $\langle Q^2\rangle\!=\!0.0088\pm 0.0001$
GeV$^2$. Analyzed with some mild model assumptions this value of $A_{\rm pv}$
has been used to extract the following result for the neutron skin thickness in
${}^{208}$Pb \cite{prexprl12}:
\begin{equation}
 \Delta r_{np}=0.33^{+0.16}_{-0.18} \;{\rm fm} \;.
\label{rnpprex1}
\end{equation}
A subsequent study has derived the result \cite{horowitz12}
\begin{equation}
 \Delta r_{np}=0.302\pm (0.175)_{\rm exp}\pm
 (0.026)_{\rm model} \pm (0.005)_{\rm strange} \;{\rm fm}\;,
\label{rnpprex2}
\end{equation}
where the last uncertainty arises from the uncertainty in the electric strange
quark form factor. Both results are much consistent with previous estimates,
although the central value is larger than the typical value reported from the
existing evidence on $\Delta r_{np}$ of ${}^{208}$Pb \cite{Tsa12}.
As mentioned, a new run PREX-II has been scheduled at the Jefferson Lab
\cite{prexii} to improve the statistics of the measurement and reach the
original 3\% accuracy in $A_{\rm pv}$.

\subsection{Theoretical considerations}

Electrons can interact with the protons and neutrons of a nucleus by exchanging
photons and $Z^0$ bosons. Whereas the photons essentially couple to protons, the
$Z^0$ bosons mostly couple to neutrons. This is due to the fact that the proton
weak charge ($Q^p_{\rm W}$) is very small in comparison with the neutron weak
charge ($Q^p_{\rm W}$). Namely, $Q^p_{\rm W} = 1 - 4 \sin^2 \theta_{\rm W}
\simeq 0.075$ (where $\theta_{\rm W}$ is the weak mixing angle) \cite{pdg12} and
$Q^n_{\rm W} \simeq -1$. The situation is practically opposite to what happens
with the proton and neutron electric charges. Therefore, high-energy elastic
electron-nucleus scattering may test not only the electric charge distribution
but also the weak charge distribution in a nucleus. The weak interaction
introduces a parity-violating term in the scattering amplitude that may be
isolated by measuring $A_{\rm pv}$. Indeed, Donnelly, Dubach and Sick
\cite{Don89} suggested almost 25 years ago that PVES on nuclei could probe the
neutron distribution via the electroweak interaction.

The parity-violating asymmetry of ultra-relativistic elastically scattered
electrons is defined as the following relative difference
\cite{prex0,prex2,vretenar00,moya10}:
\begin{equation}
A_{\rm pv} =
\frac{\displaystyle \frac{d\sigma_+}{d\Omega}-\frac{d\sigma_-}{d\Omega}}
{\displaystyle \frac{d\sigma_+}{d\Omega}+\frac{d\sigma_-}{d\Omega}} ,
\label{apv}
\end{equation}
where $d\sigma_\pm/d\Omega$ are the elastic electron-nucleus differential cross
sections for incident electrons with positive or negative helicity. Depending on
their helicity state, electrons interact with the Coulomb ($V_{\rm C}$) plus the
weak ($V_{\rm W}$) potential or with their difference:
\begin{equation}
V_\pm(r) = V_{\rm C}(r) \pm V_{\rm W}(r),
\label{potur}
\end{equation}
where the weak potential is
\begin{equation}
V_{\rm W}(r) = \frac{G_{F}}{2\sqrt{2}}\rho_{\rm W}(r),
\label{wpot}
\end{equation}
with $G_{F}$ the Fermi constant and $\rho_{\rm W}$ the weak charge density. 

We solve the scattering problem in the distorted wave Born approximation (DWBA)
\cite{roca08,Roc11,cen10} for realistic calculations of the parity violating
asymmetry defined in Eq.~(\ref{apv}). That is, we perform the exact phase shift
analysis of the Dirac equation \cite{elsepa,roca08,cen10}
\begin{equation}
[ \mbox{\boldmath$\alpha$} \!\cdot\! {\bf p}
+ V_{\pm}(r) ]\psi_{\pm} = E\psi_{\pm}
\label{dirac}
\end{equation}
for ultra-relativistic electrons. The electric and weak charge distributions of
the studied target are the main input for solving Eq.~(\ref{dirac}). To this
end, we use as a benchmark the point-like densities of protons ($\rho_p$) and
neutrons ($\rho_n$) self-consistently calculated with the nuclear mean-field
models shown in Fig.~\ref{fig2}. It is to be emphasized that such density
distributions are at the core of the nuclear energy density functionals. Thus,
in the realm of density functional theory the comparisons against experiment may
always be done directly in terms of the calculated $A_{\rm pv}$.

We obtain the electric charge density distribution $\rho_{\rm C}$ of the target
nucleus by folding the point-like proton and neutron densities with
electromagnetic proton ($G_p$) and neutron ($G_n$) form factors \cite{frie03}.
For the weak charge density distribution $\rho_{\rm W}$, we fold the
point-like densities with the electric form factors for the coupling of the
$Z^0$ boson to the proton ($G_p^{Z^0}$) and the neutron ($G_n^{Z^0}$)
\cite{cen10,prex2,moya10}:
\begin{equation}
\rho_{\rm W}(r) = \int\! d{\bf r}'
\{ 4 G_n^{Z^0}\!(r') \rho_n(|{\bf r} -{\bf r}'|)
+  4 G_p^{Z^0}\!(r') \rho_p(|{\bf r} -{\bf r}'|)\} ,
\label{rhoweak}
\end{equation}
with
\begin{eqnarray}
G_p^{Z^0}&=&+\frac{1}{4}\left(G_p-G_n\right)-G_p \sin^2\theta_{\rm W}
\label{gpz}\\
G_n^{Z^0}&=&-\frac{1}{4}\left(G_p-G_n\right)-G_n \sin^2\theta_{\rm W} , 
\label{gnz}
\end{eqnarray}
where we have neglected strange quark form factor contributions to the weak
charge distribution \cite{prex2}. Substituting $G_p^{Z^0}$ and $G_n^{Z^0}$ in
Eq.\ (\ref{rhoweak}) gives
\begin{eqnarray}
%
%
\rho_{\rm W}(r) & = & Q^p_{\rm W} \rho_{\rm C}(r) +
\nonumber \\
& & Q^n_{\rm W} \int\! d{\bf r}'
\{  G_p\!(r') \rho_n(|{\bf r} -{\bf r}'|)
+   G_n\!(r') \rho_p(|{\bf r} -{\bf r}'|) \} ;
\nonumber \\
\label{rhoweakq}
\end{eqnarray}
wherefore it follows that the weak charge density is dominated by the neutron
density of the nucleus.

Furnstahl \cite{Bro00c} showed that the neutron form factor of $^{208}$Pb at low
momentum transfer is strongly correlated with the neutron radius $r_n$ of
$^{208}$Pb in mean-field models. In fact, in a plane wave Born approximation
(PWBA) the parity-violating asymmetry is given by \cite{Don89,prex0,prex2}
\begin{equation}
A_{\rm pv}^{\rm PWBA} =
\frac{G_F q^2}{4 \pi \alpha \sqrt{2}}
\Big[4 \sin^2 \theta_W + \frac{F_n(q) - F_p(q)}{F_p(q)} \Big] ,
\label{apvpwba}
\end{equation}
where $\alpha$ is the fine structure constant and $F_n(q)$ and $F_p(q)$ are,
respectively, the neutron and proton form factors (Fourier transforms of the
point neutron and proton spatial distributions). If one takes the proton form
factor $F_p(q)$ of $^{208}$Pb at low $q$ known from experiment, the correlation
between the neutron form factor and the neutron radius of $^{208}$Pb
\cite{Bro00c} suggests, in view of Eq.~(\ref{apvpwba}), a correlation between
the parity violating asymmetry and the neutron radius of this nucleus. The
latter correlation has been shown in Ref.~\cite{ban12} in exact DWBA
calculations.

Moreover, the fact that the form factors behave as $F(q)\to 1 - q^2\langle
r^2\rangle/6$ when $q\to0$ suggests also a correlation between the parity
violating asymmetry and the neutron skin thickness of the target nucleus in the
low-momentum transfer regime, since for $q\to0$ one has
\begin{equation}
A_{\rm pv}^{\rm PWBA} = \frac{G_F q^2}{4\pi \alpha \sqrt{2}}
\Big[ 4 \sin^2 \theta_W - \Delta r_{np} (r_n+r_p) \frac{q^2}{6}
+ \cdots  \Big] .
\label{apvpwba1}
\end{equation}
Here we have written $r_n^2-r_p^2$ as $\Delta r_{np} (r_n+r_p)$. For realistic
results and comparisons against experiment, however, one cannot use the simple
and insightful PWBA and full DWBA calculations have to be performed. Note that
the Coulomb distortions can correct the PWBA results by more than 40\% of the
value found in the PWBA.

\subsection{Discussion of results for the parity violating asymmetry
in $^{208}$Pb}
\label{resultsI}

\begin{figure}[t]
\includegraphics[width=1.0\columnwidth,clip=true]{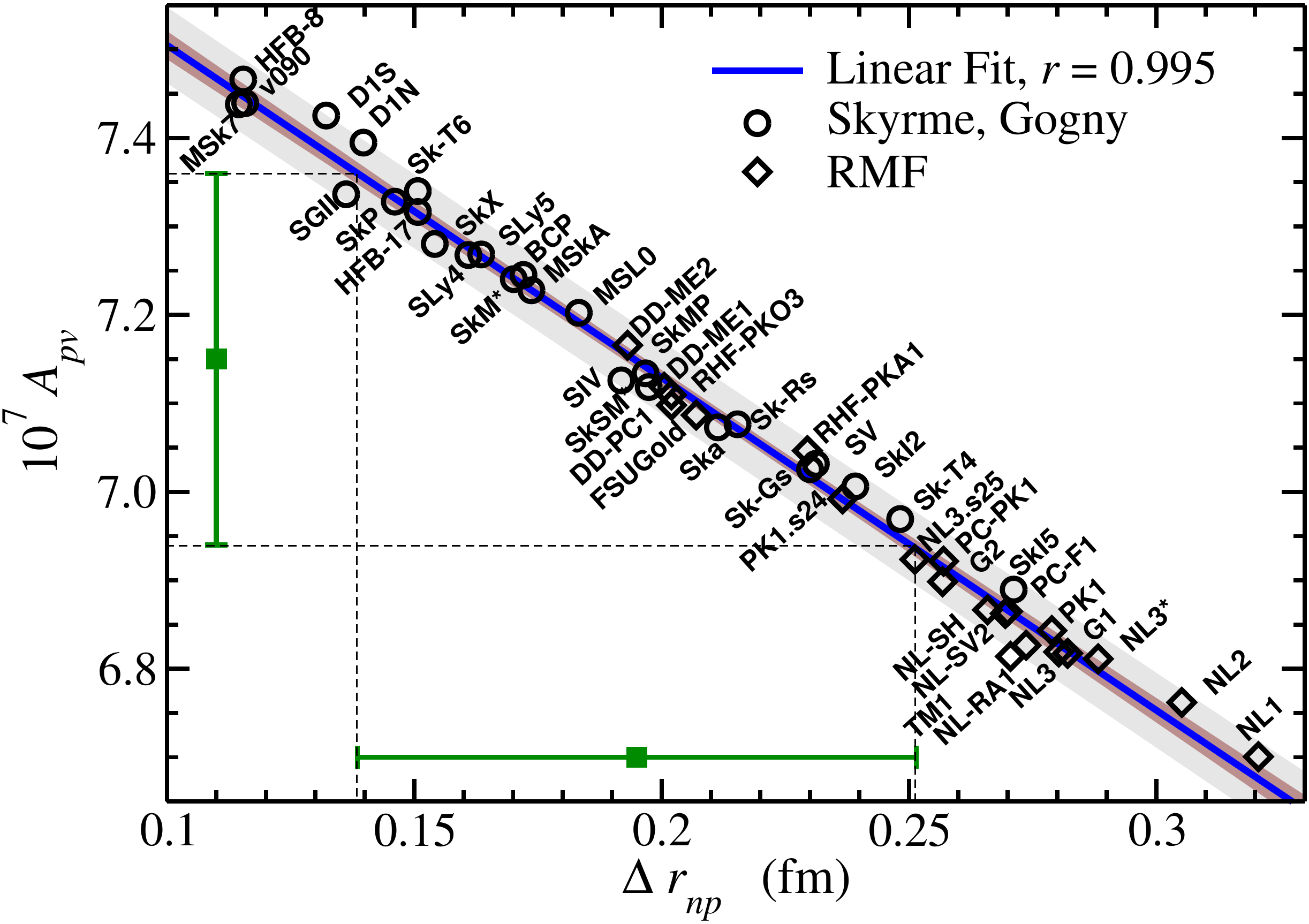}
\caption{\label{fig7} (Color online) Predictions of nuclear mean-field models
for the parity violating asymmetry as a function of the neutron skin thickness
in ${}^{208}$Pb. The results have a linear correlation coefficient $r=0.995$. A
linear fit is shown yielding $A_{\rm PV}$(ppm)$= 0.788 - 0.375 \Delta
r_{np}$(fm). For the purpose of illustration of the accuracy in $\Delta r_{np}$
that a 3\%-accuracy value of $A_{\rm pv}$ would imply, an arbitrarily chosen
central point for $A_{\rm pv}$ with 3\% error bars is plotted. The thinner and
thicker shadowed regions represent, respectively, the 95\%-confidence band and
95\%-prediction band of the linear regression (see text). Figure adapted from
Ref.~\cite{Roc11}.}
\end{figure}

We display in Fig.~\ref{fig7} the relation between the parity violating
asymmetry $A_{\rm pv}$ and the neutron skin thickness $\Delta r_{np}$ in
${}^{208}$Pb within the realm of nuclear energy density functionals
\cite{Roc11}. The scattering problem has been solved in exact DWBA calculations
for electrons that scatter from lead at an energy of $1.06$ GeV and an angle of
$5^\circ$, which is close to the operating conditions of PREX.

The corpus of 47 nuclear models considered in Fig.~\ref{fig7} is representative
of models of very different kind. We have used non-relativistic Skyrme forces
(all models of Fig.~\ref{fig7} that start by S and the models HFB-8, HFB-17,
v090, MSk7, MSkA, and MSL0), Gogny forces (models D1S and D1N), and the
finite-range BCP functional. On the other hand, we have considered relativistic
mean-field models based on effective Lagrangians. In this case, in
Fig.~\ref{fig7} we have included RMF models with non-linear meson couplings
(these are the models with names starting by NL or PK, plus the models FSUGold,
G1, G2, and TM1), meson-exchange models with density-dependent couplings (DD-ME
and RHF-PK models), and finally, zero-range point-coupling models (DD-PC1,
PC-PK1, and PC-PF1 models). The original references to these nuclear models can
be found in the papers \cite{Cen09,cen10,carbone10,chen10,baldo08}.

All of the models that we have allowed in our analysis predict the {\sl charge}
radius of ${}^{208}$Pb to better than 1\% with respect to the experimental value
(it is the same accuracy level as the 1\% goal of PREX for the neutron radius).
We have not either included models that fail significantly to
predict the observed binding energy of $^{208}$Pb. It would be less realistic
for constraining the unknown neutron radius of $^{208}$Pb to include results for
$A_{\rm pv}$ calculated with the neutron densities predicted by
models that fail to describe the well-established properties of $^{208}$Pb, like
its charge radius (one has $r_{\rm ch} = 5.5012(13)$ fm in $^{208}$Pb
\cite{angeli13}). Let us remark also that we have not allowed more than two
models of the same kind fitted by the same group and protocol in order to
minimize potential biass in our conclusions.

It can be seen from Fig.~\ref{fig7} that the nuclear mean-field models predict a
highly linear relation between $A_{\rm pv}$ and $\Delta r_{np}$ in ${}^{208}$Pb,
in spite of the different theoretical grounds that underly the large variety of
considered models. A linear correlation coefficient $r=0.995$ is found. We also
plot in Fig.~\ref{fig7} the 95\%-confidence band of the linear regression
(colored thin inner band). The so-called confidence band represents the boundary
of all of the possible straight lines fitting the results within a
95\%-confidence level, see for example Ref.~\cite{regres} for details. The fact
that this band is very thin in Fig.~\ref{fig7} is another indicator of the
strong linear correlation of the results. Also displayed in Fig.~\ref{fig7} is
the so-called prediction band \cite{regres} at a 95\%-confidence level (it is
the wider colored band that basically coincides with the envelope of the models
in the figure).

We next explore the accuracy in the determination of $\Delta r_{np}$ that can be
obtained from a 3\% accuracy in $A_{\rm pv}$ (as proposed in PREX-II
\cite{prexii}) assuming the correlation predicted by the nuclear energy density
functionals. For illustration, we have placed in Fig.~\ref{fig7}
an arbitrary value of $A_{\rm pv}$ (viz., 0.715 ppm) with an attached 3\% error
bar (viz., $\pm 0.022$ ppm) as depicted in the figure. Given that the
correlation between $A_{\rm PV}$ and $\Delta r_{np}$ is highly linear, the
conclusions about the resulting accuracy in $\Delta r_{np}$ will be very similar
for other $A_{\rm pv}$ values, at least if they are not very
far away. Even though our test central value $A_{\rm pv}=0.715$ ppm is not the
one measured in the first run of PREX ($A_{\rm pv}=0.656\pm(0.060)_{\rm stat}
\pm (0.014)_{\rm syst}$ ppm), they agree to within the uncertainties.

In Fig.~\ref{fig7}, a value $A_{\rm pv}=0.715$ ppm corresponds to a point
located
around the central region of the theoretical predictions for $A_{\rm pv}$ in the
considered models. It implies a central value for $\Delta r_{np}$ that lies both
within the range between 0.15 and 0.22 fm suggested by the experiments with
strong probes \cite{zen10,cen10}, and within the range between 0.14 and 0.20
fm suggested by recent constraints on the nuclear EOS derived from observed
masses and radii of neutron stars \cite{stei10,hebe10}. While the
latter constraints and a majority of studies point toward a fairly thin neutron
skin in lead \cite{Tsa12}, it is to be mentioned that ruling out a
thick neutron skin as suggested by the central value of the first run of PREX
\cite{prexprl12,horowitz12} could be premature~\cite{fattoyev2013}.

The 3\%-accuracy test value of $A_{\rm pv}$ plotted in Fig.~\ref{fig7} implies
through the precise universal correlation displayed in the figure a neutron skin
thickness $\Delta r_{np}=0.195 \pm 0.057$ fm for $^{208}$Pb.
Knowing $\Delta r_{np}$ and the proton radius of $^{208}$Pb by unfolding the
finite size of the proton charge from the precisely known charge radius of this
nucleus \cite{vries87,angeli13}, the neutron radius $r_n$ can be obtained.
One can assume for $r_n$ the same uncertainty of $\Delta r_{np}$ ($\pm 0.057$ fm
in our test example) because the accurate charge radius of $^{208}$Pb
\cite{angeli13} has a negligibly small uncertainty in comparison with $\Delta
r_{np}$ and $r_n$. Therefore, the projected accuracy in $r_n$ is better than
$\pm 0.06$ fm. This implies an accuracy of about 1\% in $r_n$, which
is in total support of the goal of the PREX experiment. It may be underscored
that the current analysis allows one to predict the NST and the neutron
radius without specific assumptions about the shape of the neutron and proton
density profiles.

In Ref.~\cite{Roc11} we also studied directly the correlation between the values
of $A_{\rm pv}$ and of $r_n$ predicted in $^{208}$Pb by the mean-field models.
For the same models of Fig.~\ref{fig7}, the calculated results of $A_{\rm pv}$
and $r_n$ showed a linear trend with a correlation coefficient $r=0.974$. That
is, the correlation of $A_{\rm pv}$ with $r_n$ is a little less precise than the
correlation of $A_{\rm pv}$ with $\Delta r_{np}$. If we repeat the test that we
have done before for $\Delta r_{np}$, a 3\%-accuracy value of $A_{\rm pv}$
placed at 0.715 ppm projects into a neutron radius $r_n$ of $5.644\pm0.065$ fm
for $^{208}$Pb \cite{Roc11} (the experiments involving strong probes suggest a
range of roughly 5.52 fm to 5.67 fm for the central value of $r_n$). Since the
correlation of $A_{\rm pv}$ with $r_n$ is not as strong as the correlation of
$A_{\rm pv}$ with $\Delta r_{np}$, for the 3\% accuracy assumed in $A_{\rm pv}$
a little higher uncertainty is derived from the $A_{\rm pv}$ vs $r_n$
correlation than from the $A_{\rm pv}$ vs $\Delta r_{np}$ correlation (i.e,
$\pm0.065$ fm uncertainty instead of $\pm0.057$ fm uncertainty, or a 14\% more).

\subsection{Constraining the slope of the density dependence of the symmetry
energy at saturation from PVES}

\begin{figure}[t]
\includegraphics[width=1.0\columnwidth,clip=true]{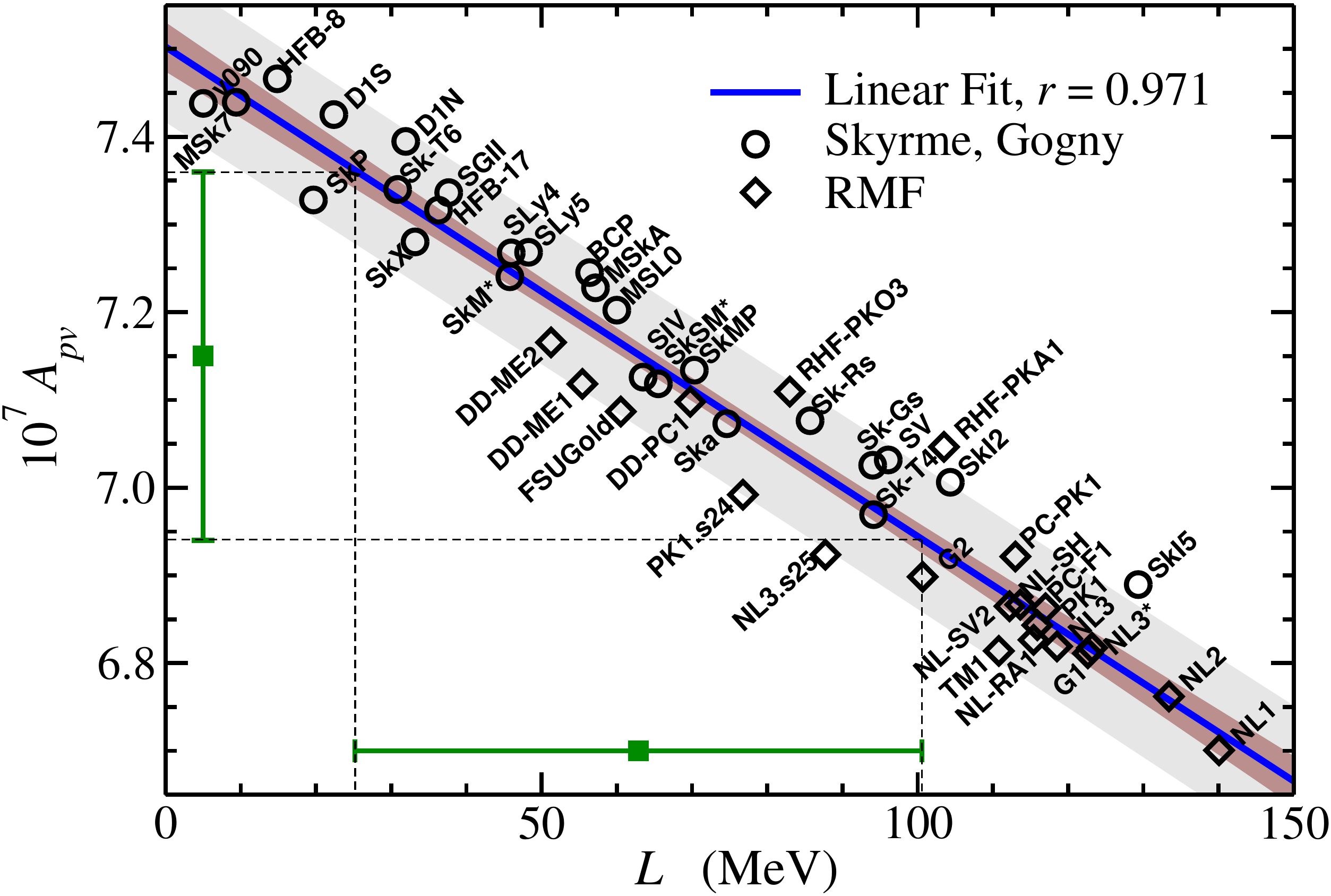}
\caption{\label{fig8} (Color online) Predictions of nuclear mean-field models
for the parity violating asymmetry in ${}^{208}$Pb as a function of the slope
parameter of the nuclear symmetry energy at saturation density. The results have
a linear correlation coefficient $r=0.971$. A linear fit is shown yielding
$A_{\rm PV}$(ppm)$= 0.750 - 0.00055 L$(MeV). For the purpose of illustration of
the accuracy in $L$ that a 3\%-accuracy value of $A_{\rm pv}$ would imply, an
arbitrarily chosen central point for $A_{\rm pv}$ with 3\% error bars is
plotted. The thinner and thicker shadowed regions represent, respectively, the
95\%-confidence band and 95\%-prediction band of the linear regression (see
text).}
\end{figure}

The determination of the density dependence of the nuclear symmetry energy has
been identified as one of the most outstanding questions in nuclear physics for
its deep implications for a wide variety of problems and phenomena. From the
results discussed in the previous subsection and the fact that $\Delta r_{np}$
of a heavy nucleus is strongly correlated with the slope $L$ of the symmetry
energy at the saturation density (see Fig.~\ref{fig2}), we can expect a linear
trend of $A_{\rm pv}$ with respect to the $L$ parameter in nuclear mean-field
models. This expectation is confirmed by the plot shown in Fig.~\ref{fig8}. It
implies that a measurement of the parity violating asymmetry in ${}^{208}$Pb
directly tests the density dependence of the nuclear symmetry energy.

Note that the models with a stiffer symmetry energy around saturation density
(models with larger $L$ values) have, consequently, a smaller symmetry energy at
the subsaturation densities which are relevant for finite nuclei. Such models
produce thicker neutron skins, and have smaller values of the parity violating
asymmetry as it can be seen from the negative slope of $A_{\rm pv}$ as a
function of $L$ in Fig.~\ref{fig8}.

The linear correlation coefficient of the results for $A_{\rm pv}$ in
${}^{208}$Pb and $L$ is again quite high ($r=0.971$). From the confrontation of
$A_{\rm pv}$ with $L$ we find the following linear relation between these two
quantities in nuclear models:
\begin{equation}
A_{\rm pv} \;{\rm (ppm)} \;=\; 0.750 - 0.00055 L \;{\rm (MeV)} \;.
\label{fitalrL}
\end{equation}
One of the main merits of an electroweak extraction of the critical parameter
$L$ is that it would be mostly free from strong interaction uncertainties.

Assuming as before a 3\%-accuracy value of $A_{\rm pv}$, placed arbitrarily at
0.715 ppm, one finds $L=64\pm 39$ MeV. This means that a 3\%-accuracy in $A_{\rm
pv}$ implies roughly a window of $\pm 40$ MeV in $L$. Though the central value
of $L$ depends on the assumption of 0.715 ppm, the spread in the determination
of $L$ associated with a 3\%-accuracy measurement of $A_{\rm pv}$ almost does
not. Hence, one is bound to conclude that a 3\% accuracy in the parity violating
asymmetry does not impose a very narrow constraint on $L$. However, the
constraint can be of great interest because PREX for the first time probes
neutrons in a heavy nucleus with an electroweak probe instead of less clean
strong probes.

Let us mention that the high linearity of the correlation between between
$A_{\rm pv}$ and $L$ (and between $A_{\rm pv}$ and $\Delta r_{np}$) found in our
theoretical study shows that if $A_{\rm pv}$ could be measured with $\leq\! 1$\%
accuracy, then $L$ could be estimated up to within a narrow window close to $\pm
10$~MeV accuracy (and $\Delta r_{np}$ within a window of about $\pm 0.02$ fm).
Such a precise determination of $L$ through an electroweak measurement could
have lasting impact in diverse areas of nuclear physics and astrophysics. Beyond
doubt, a $\leq\! 1$\% accuracy in $A_{\rm pv}$ represents a formidable
experimental challenge. Hopefully, the PREX-II experiment \cite{prexii} and new
facilities such as MESA in Mainz~\cite{sfie13} will pave the way in a near
future for such a goal.

\section{Size and shape of the neutron distribution in ${}^{208}$Pb and symmetry
energy}
\label{resultsII}
 
The extraction of the neutron skin thickness from experiment does not indicate
in an unambiguous way whether the neutron skin is caused by an extended bulk
radius, by a larger surface width of the neutron density, or by a combination of
both effects \cite{Trz01,Swi05}. This is a relevant aspect in experimental
characterizations of neutron
densities~\cite{Trz01,Jas04,fried03,fried05a,fried05b}. It is also an
interesting aspect from a theoretical point of view because in $^{208}$Pb it is
related with the density dependence of the symmetry energy~\cite{cen10}.

The analysis of data in different types of experiments (such as electron
scattering,  nucleon scattering, experiments in exotic antiprotonic and pionic
atoms, etc.) often involves parametrized nucleon densities. A popular profile is
the two-parameter Fermi (2pF) function
\begin{equation} 
\label{2pF}
\rho(r) = \frac{\rho_0}{1+ \exp{ [(r-C)/a] }} ,
\end{equation} 
where $\rho_0$ is central density, $C$ is the half-density radius, and $a$ is
the diffuseness parameter. Albeit simple, a 2pF profile encodes the two most
important quantities that characterize the shape of the density distributions,
namely the position and thickness of the nuclear surface.

In the literature it has been found useful to distinguish two limiting
situations in the formation of a neutron skin in a nucleus assuming 2pF shapes
\cite{Trz01,Jas04,fried03,fried05a,fried05b}. On the one hand, in the so-called
{\sl halo-type} density distributions the 2pF nucleon profiles have $C_n-C_p=0$
and $a_n-a_p>0$. On the other hand, in the {\sl skin-type} density distributions
the 2pF nucleon profiles have $a_n-a_p=0$ and $C_n-C_p >0$.

We can analyze whether the theoretical nuclear mean-field models support either
(halo or skin) type of density distributions in $^{208}$Pb by applying the
strategy of Refs.\ \cite{warda10,cen10}. That is, in order to obtain the
predictions of the mean-field models for the central radius and diffuseness of
the nucleon densities, we fit 2pF functions to the self-consistently calculated
nucleon densities of the various models studied in the previous section. There
is no unique parametrization of a given density profile with a 2pF function.
Following Refs.\ \cite{warda10,cen10}, we fit the parameters of Eq.\ (\ref{2pF})
to reproduce the zeroth, second, and fourth moments of the actual
self-consistent mean-field density distributions. This method has been proven to
fit accurately the surface region of any realistic density profile given as an
input \cite{warda10,cen10}.

\begin{figure}[t]
\includegraphics[width=1.0\columnwidth,clip=true]{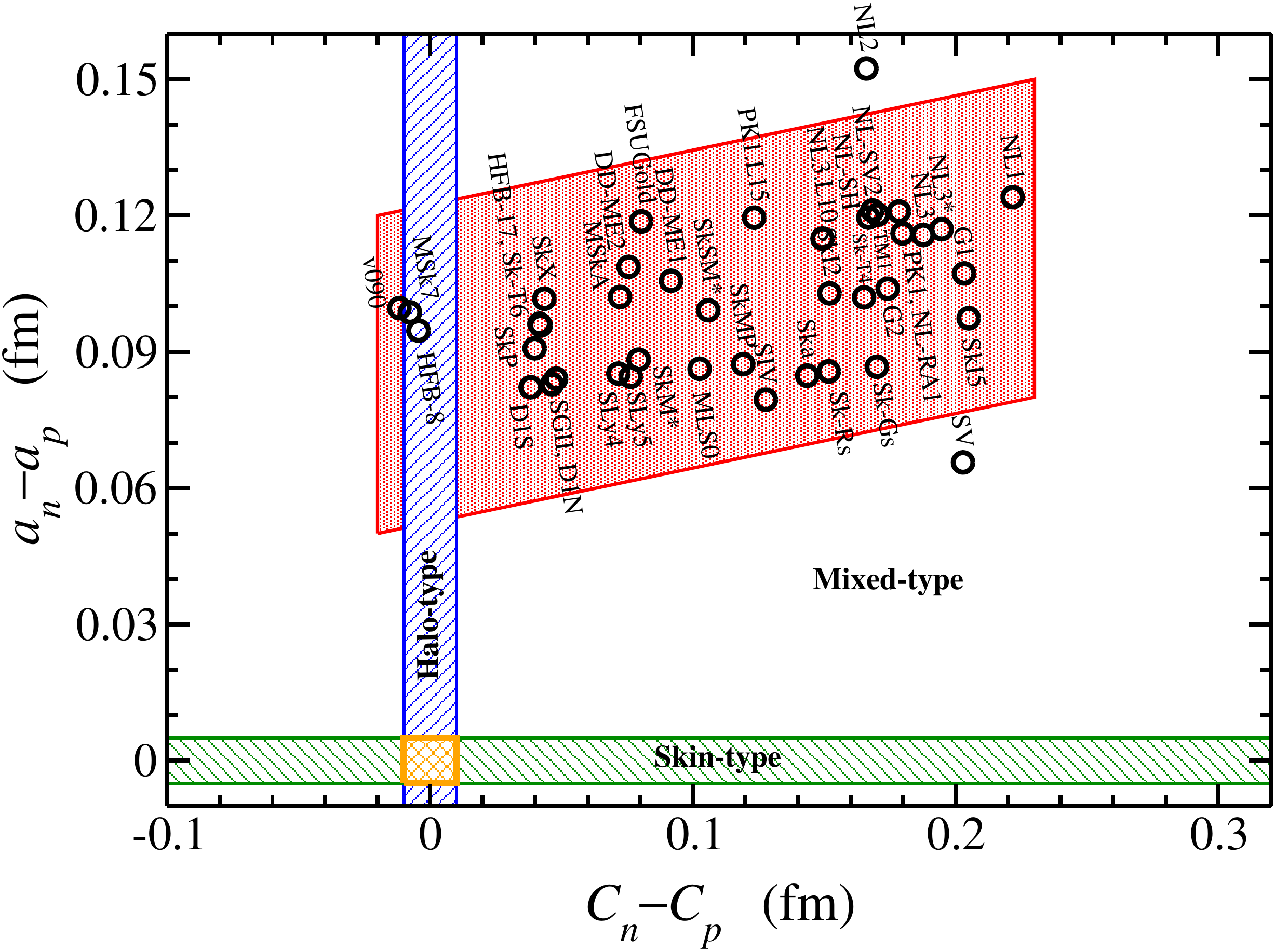}
\caption{\label{fig9} (Color online) Difference $a_n-a_p$ between the surface
diffuseness of the neutron and proton density profiles in ${}^{208}$Pb against
the corresponding difference $C_n-C_p$ of the central radii as predicted by
the employed mean-field models.}
\end{figure}

We display in Fig.~\ref{fig9} the difference $a_n-a_p$ between the surface
diffuseness of neutrons and protons against the difference $C_n-C_p$ between the
half-density radii for $^{208}$Pb in the present nuclear models. It is seen that
models with a stiffer symmetry energy around saturation (i.e., models with
larger values of the $L$ parameter) have larger values of the $C_n-C_p$
difference. The difference $a_n-a_p$, however, is rather insensitive to the
stiffness or softness of the symmetry energy \cite{cen10}.

Inspection of Fig.~\ref{fig9} shows that the nuclear energy
density functionals overall predict density distributions with a mixed character
between the halo and skin types. Nevertheless, it is to be observed that the
models
where the symmetry energy is softer do favor the halo-type densities, like the
case of the HFB-8, MSk7, and v090 interactions. In contrast, we can see in
Fig.~\ref{fig9} that no model predicts a skin-type distribution for $^{208}$Pb,
as the difference $a_n-a_p$ is in all cases systematically away from a vanishing
value. It may be reminded that in Refs.\ \cite{Trz01,Jas04} the experimental
extraction of neutron skins in antiprotonic atoms assumed halo-type density
distributions. The theoretical predictions derived from nuclear mean-field
models, thus, suggest that such an assumption is more plausible (especially if
the symmetry energy is soft) than assuming that the neutron skin is formed by
the mechanism of skin-type density distributions where $a_n-a_p=0$.

The specific values of the $C_n$ and $a_n$ parameters of the 2pF functions
describing the neutron mean-field densities of $^{208}$Pb can be found in Table
III of Ref.~\cite{cen10} for many of the nuclear interactions we have considered
here. Summarizing our results, we find that the neutron half-density radius
$C_n$ spans a range between approximately 6.6 and 7.0 fm. Meanwhile, the neutron
surface diffuseness $a_n$ lies in a window that spans at most from 0.5 to 0.6
fm; indeed, the majority of considered nuclear mean-field models predict $a_n$
within 0.545 and 0.565 fm.

As mentioned in the previous section where we discussed parity-violating elastic
electron scattering in $^{208}$Pb, the parity violating asymmetry $A_{\rm pv}$
in the PREX experiment is obtained at one angle. Hence, assuming the electric
charge distribution of $^{208}$Pb known from experiment, a determination of the
shape of the neutron density of $^{208}$Pb parametrized by a 2pF profile
requires a second measurement of $A_{\rm pv}$ at another angle \cite{ban12}.
However, we still may get some information from a single determination of the
parity violating asymmetry if we describe the neutron density profile by
assuming a 2pF function and restrict the neutron surface diffuseness to lie
within the window predicted by the nuclear mean-field models.

Working with a 2pF function, the neutron radius can be accurately calculated
using the following formula:
\begin{equation} 
\label{EQUATION_A2}
\langle r^2\rangle_n^{1/2} = \sqrt{\frac{3}{5}} \, C_n
 \left[ 1+\frac{7\pi^2}{3} \frac{a_n^2}{C_n^2} \right]^{1/2}
\end{equation} 
(see, e.g., the appendix of Ref.\ \cite{warda10}). For any given value of the
neutron radius, if we vary $a_n$ within a prescribed range of values, it will
produce a family of 2pF functions having the same neutron radius and different
central radii $C_n$. That is, we obtain a family of different neutron shapes
having the same rms radius. We can compute the parity violating asymmetry for
that family of 2pF neutron profiles, and then repeat the same exercise with a
new value of the neutron radius.

\begin{figure}[t]
\includegraphics[width=1.0\columnwidth,clip=true]{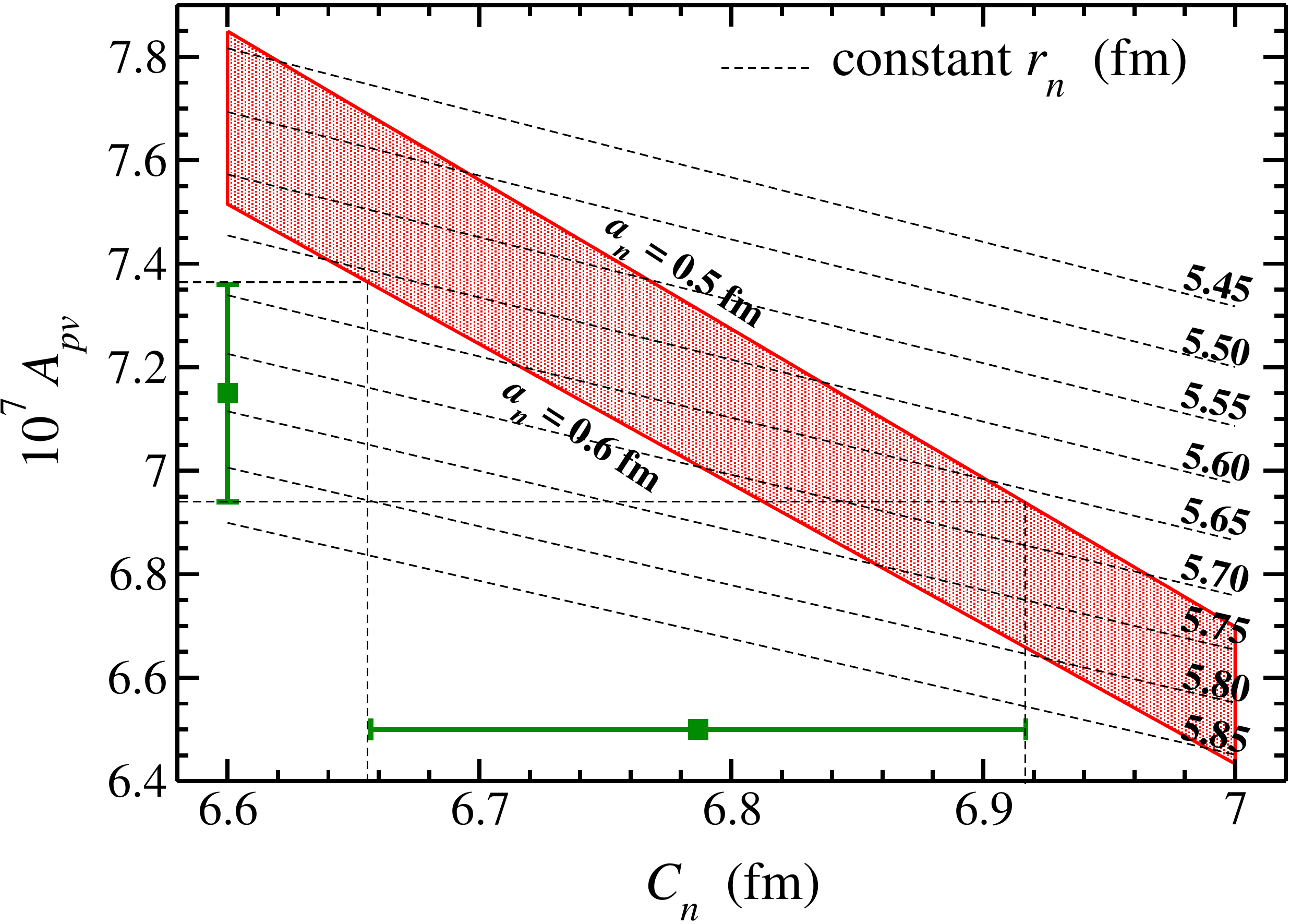}
\caption{\label{fig10} (Color online) Parity violating asymmetry in ${}^{208}$Pb calculated using 2pF neutron densities and the experimental charge density of Ref.~\cite{vries87}. The results are plotted as a function of the half-density radius $C_n$ of the 2pF profile. Lines of constant neutron radius are also drawn. The neutron surface diffuseness has been constrained to a range $0.5~{\rm fm} \leq a_n \leq 0.6$ fm. For the purpose of illustration of the accuracy in $C_n$ that a 3\%-accuracy value of $A_{\rm pv}$ would imply, an arbitrarily chosen central point for $A_{\rm pv}$ with 3\% error bars is plotted.}
\end{figure}

We display in Fig.~\ref{fig10} the calculated parity violating asymmetry by
employing the experimental charge density of Ref.~\cite{vries87} and 2pF neutron
densities with the mentioned constraints. The results are plotted as a function
of $C_n$ of the 2pF neutron profiles. Note that in Fig.~\ref{fig10} we are not
using any densities of mean-field models. (For a calculation of $A_{\rm pv}$
with 2pF neutron and proton profiles with the parameters fitted to the
theoretical mean-field densities, see section III.D of Ref.~\cite{cen10}.)
We also show in Fig.~\ref{fig10} the surfaces of constant
neutron radius projected in the $A_{\rm pv}$-vs-$C_n$ plane. As expected, we
find a linear correlation between $A_{\rm pv}$ and $C_n$ since the central
radius gives the main contribution to the rms radius of a 2pF function in
${}^{208}$Pb (cf.\ Eq.\ \ref{EQUATION_A2} and Ref.~\cite{cen10}).

In analogy to our previous analysis in Section \ref{secpves}, we set on top of
Fig.~\ref{fig10} a test value of $A_{\rm pv}$ with a 3\% error bar, placed at
$0.715$ ppm as we did before in Section \ref{secpves}. From such a test
constraint and a neutron surface diffuseness
$a_n=0.55\pm 0.05$ fm, we find a half-density radius $C_n=6.79\pm 0.13$ fm for
the neutron density profile of ${}^{208}$Pb. This implies a neutron rms radius
$r_n=5.643\pm 0.115$ fm (and a NST $\Delta r_{np}=0.201\pm
0.115$ fm). Therefore, a 3\% uncertainty in $A_{\rm pv}$ is projected into a 2\%
uncertainty in the neutron radius through the present analysis with 2pF neutron
shapes. We note that the central values of the current predictions closely
coincide with the values extracted in our analysis carried out in Section
\ref{resultsI} (namely, $r_n=5.644\pm 0.065$ fm and $\Delta
r_{np}=0.195 \pm 0.057$ fm). However, the expected uncertainty is now about
twice as large. This uncertainty could be narrowed down if we had not been so
conservative in our estimate of $a_n$ and, instead of assuming $a_n=0.55\pm
0.05$ fm, we had taken $a_n$ between 0.545 and 0.565 fm which is the prediction
of a majority of mean-field models~\cite{cen10}.

\section{Discussion of results and comparison with other estimates of the slope of the symmetry energy at saturation}
\label{sec:2.4}

During the last years a major effort has been made in extracting the
density dependence of the symmetry energy from experimental measurements and
theoretical studies of very different nature. In Fig.\ \ref{fig1} of the
Introduction we have depicted a sample of such methods and predictions for
the slope parameter $L$. Here we aim at a short presentation of some of these
studies and to facilitate a comparison of existing $L$ constraints (we emphasize
that it would be near to impossible to mention in this section all
determinations and we apologize for those studies that inadvertently may be
absent below). Additional information can be found in the quoted references and
in \cite{Tsa12,Lat13}. For detailed accounts of several studies mentioned
in the following, see the related contributions to the present volume.

On the one side, there exist sophisticated microscopic calculations of the
neutron matter EOS based on chiral effective field theory that
combined with observed neutron star masses and radii \cite{hebe10} predict $L
\sim 30$--60 MeV. A determination of the EOS of neutron matter from quantum
Monte Carlo calculations including realistic two- and three-nucleon
forces \cite{Gan10} is compatible with similar $L$ values.
The results from microscopic Brueckner-Hartree-Fock calculations including
effective three-body forces give $L=66.5$ MeV \cite{vid11}.
These values for the slope of the symmetry energy are consistent
with the range around $L\sim40$--60 MeV predicted by an empirical EOS
constrained exclusively on the basis of astrophysical observations (namely, a
heterogeneous data set of six neutron stars) \cite{stei10}.

Studies of heavy-ion collisions of neutron-rich nuclei at intermediate energies
are powerful sources of information to constrain $L$ \cite{Bar05,Li08}.
Sometimes modeled forms of the symmetry energy, such as Eq.~(\ref{eq11a}), are
implied to deduce $L$. Simulations of isospin diffusion data with isospin- and
momentum-dependent transport models \cite{Tsa04,Ste05a,Li05a,Li05b,Li08} have
allowed to estimate $L$ in a range between 50 and 110 MeV. Improved quantum
molecular dynamics simulations describing both isospin diffusion data and double
ratios of neutron and proton spectra, support a similar range of $L$ values for
a symmetry energy at saturation between 28 and 34 MeV \cite{Tsa09}. Experiments
on transverse collective flows of hydrogen and helium isotopes
\cite{Koh10,Koh11} favor similar values of $L$, as well as the predictions from
neutron-proton emission data \cite{Fam06} ($L\sim 50$ MeV) and the
isotopic-scaling results of Ref.~\cite{She07} ($L\sim 65$ MeV).

Another source of information comes from the study of binding energies, analog
states, and neutron skins. The study of masses
\cite{TF94a,TF94b,Dan03,Liu10,mol12,Dong2012,Agr12} leads to $L$ values between
roughly 50 and 85 MeV. In particular, the new finite-range droplet model (FRDM)
predicts $L=70\pm15$ MeV \cite{mol12}. Similar values of $L$ as from nuclear
masses are supported by studies of $\alpha$-decay energies ($L=61\pm22$ MeV)
\cite{Dong2013}, Fermi-energy differences in nuclei ($L=47\pm18$ MeV)
\cite{Wang13a}, and isospin effects in nuclear charge radii ($L=54\pm19$ MeV)
\cite{Wang13b}. Excitation energies of isobaric analog states
\cite{Dan09} point to a range $L \sim 78$--111 MeV. Information on neutron
distributions and skins can be obtained from nucleon elastic scattering.
Recently, the neutron density of lead has been measured by polarized
proton elastic scattering \cite{zen10} giving $\Delta r_{np} =
0.211^{+0.054}_{-0.063}$ fm in ${}^{208}$Pb, which suggests $L= 30$--110
MeV through the $\Delta r_{np}$--$L$ correlation (cf.\ Fig.~\ref{fig2}). An
estimate of the NST in many nuclei is provided by the information obtained from
antiprotonic atoms \cite{Trz01,Jas04,Klo07} discussed in Sections
\ref{sec:2.2} and \ref{sec:2.3}. Using the neutron skins of 26 stable
nuclei found from this technique, we predicted a range $L= 30$--80~MeV for the
slope of the symmetry energy. The study of existing neutron skin data in Sn
isotopes taken in combination with constraints from heavy-ion collisions leads
to $L= 58\pm18$ MeV \cite{chen10}. Using the data on the NST in Sn isotopes and
isotope binding energy differences, a range $L= 45.2\pm10$ MeV has been obtained
\cite{Zhang2013}. Most of the mentioned results are seen to be in good
agreement among them and also with the value $L=52.7$ MeV extracted from global
nucleon optical potentials \cite{Xu2010}.

In order to measure neutron radii and skins, as we have discussed in the
previous sections, parity-violating elastic electron scattering is a technique
free from most strong interaction uncertainties
\cite{Don89,prex0,prex2,vretenar00,moya10}. The feasibility of this type of
experiments in heavy nuclei has been recently demonstrated by the PREX
collaboration \cite{prexprl12}. The first measurement \cite{prexprl12}
had a limited statistics and the analysis of the experimental data is compatible
with a wide range of $L$ values (cf.\ Fig.~\ref{fig1}). A better accuracy is 
pursued in a future experiment PREX-II \cite{prexii}.

Isovector giant resonances are a further experimental source of information on
the symmetry energy. Indeed, in Ref. \cite{Tri08} the study of the Giant
Dipole Resonance (GDR) in ${}^{208}$Pb was
used to constrain the value of the symmetry energy at $\rho=0.1$ fm$^{-3}$ to be
within 23.3 and 24.9 MeV. A study on the Giant Quadrupole Resonance (GQR) in
${}^{208}$Pb provides a consistent constraint $S(\rho=0.1$ fm$^{-3})=23.3\pm0.6$
MeV \cite{Roca13}. Although the nature of the low-lying dipole
strength ---the so-called Pygmy Dipole Resonance (PDR)--- is still under debate,
it has been identified by some authors as one of the observables that can shed
light into a better characterization of
the properties of the nuclear symmetry energy around saturation. In
Ref.~\cite{Kli07}, the measured low-energy dipole response in neutron-rich Sn
and Sb isotopes and available data on ${}^{208}$Pb allowed to derive
$J=32.0\pm 1.8$ MeV and a pressure of pure neutron matter at saturation
$p_0=2.3\pm 0.8$ MeV fm$^{-3}$, implying $L=43\pm 15$ MeV. The same
observable was studied in $^{68}$Ni and $^{132}$Sn in Ref.~\cite{carbone10} and
$L=64.8\pm15.7$ MeV was obtained.
More recently, experiments using inelastic scattering of polarized protons on
$^{208}$Pb at very forward angles \cite{Tam11} combined with previous
measurements on different energy ranges have provided the complete electric
dipole response in $^{208}$Pb with high accuracy. A theoretical analysis of the
experimental data \cite{Pie12} leads to $\Delta r_{np} = 0.168\pm 0.022$ fm,
which implies $L \sim 25$--60 MeV based on the
strong correlation shown in Fig.~\ref{fig2}. A more recent theoretical investigation
has found a relation between the $L$ and $J$ parameters once the
experimental value of the polarizability in a heavy nucleus is known. In
$^{208}$Pb, such a relation is 
$L\!=\!-146\pm(1)_{\rm theo.} \!+\!\big[6.11\pm(0.18)_{\rm exp.}\pm(0.26)_{\rm
theo.} \big]J$. Assuming a plausible range for $J$ of 29--33 MeV
\cite{Tsa12,Lat13}, the constraint $L = 43 \pm 16$ MeV has been
obtained~\cite{Roca13a}.

Concluding, all the mentioned techniques predict values of the slope of the
symmetry energy at saturation $L$ lying in a range between about 20 and 130 MeV.
In spite of the discrepancies in the details, the various findings from
isospin-sensitive observables, including our study of neutron skins, seem to
suggest a rather soft nuclear symmetry energy around saturation. Specifically, a
weighted average of the different estimates on $L$ shown in Fig.~\ref{fig1}
---neglecting the value reported in Ref.~\cite{Fam06} since no error bar was
available--- suggests a rather narrow compatible window for the {\sl central}
value of the $L$ parameter that lies in the range $\sim 44-68$ MeV, though the
large uncertainties indicate a standard deviation of about 50 MeV.

\section{Summary and outlook}
\label{summary}

We have reviewed our main findings on the density content of the symmetry energy and the neutron skin thickness $\Delta r_{np}$ in a heavy nucleus such as $^{208}$Pb reported in earlier literature \cite{Cen09,War09,cen10,Roc11}. We have discussed possible extractions of such magnitudes from two experimental techniques, one involving the strong interaction and the other one involving the electro-weak interaction.

First, we have used insights from the DM \cite{MSa,MSb,MSc,Mye80} to study the
NST of nuclei. In the DM picture the neutron skin consists of two contributions.
One of them comes from the separation between the neutron and proton mean
surfaces and the other one from the fact that the surface widths of the neutron
and proton densities are different. In the DM, $\Delta r_{np}$ is basically
driven by the ratio between the symmetry energy at saturation $J$ and the
surface stiffness coefficient $Q$. This ratio $J/Q$ is directly related to the
symmetry energy coefficient of a finite nucleus \cite{MSa,MSb,MSc}.
To relate the NST in a nucleus with the density content of the symmetry
energy in the EOS, use has been made of a generic relation in mean-field models,
where the DM symmetry energy coefficient in a nucleus and the symmetry
energy in nuclear matter at a subsaturation density $\rho$ nearly coincide. This
relation allows one to recast the expression for $\Delta r_{np}$ of the DM in
terms of the slope parameter of the symmetry energy $L$. The result provides a
clear-cut insight into the correlation, first established by Brown
\cite{Bro00a,Bro00b}, between the NST in $^{208}$Pb and the
slope of the symmetry energy at saturation.

The neutron skins of 26 nuclei from $^{40}$Ca to $^{238}$U measured in
antiprotonic atoms \cite{Trz01,Jas04} are the largest set of skins extracted
uniformly with the same experimental technique for stable isotopes along the
periodic table. These
data show an average linear behavior with $I$ \cite{Trz01,Jas04}. From the
fit of the DM formula to the skin data of Refs.\ \cite{Trz01,Jas04}, we found
that the ratio $J/Q$ lies in a range of 0.6--0.9 and the slope of the symmetry
energy $L$ is between 30 and 80 MeV. Though the neutron skins derived
from antiprotonic atoms are to some extent model dependent and have for some
nuclei large uncertainties, we find that the deduced estimate of the density
dependence of the symmetry energy is very compatible with other analysis. In
particular, it has a large overlap with the $L$ predictions from distinct
experimental information such as heavy-ion collisions, proton-nucleus
scattering, and giant resonances in neutron-rich nuclei. Note that the
extraction of almost all of these estimates is largely ruled by
the strong interaction and, therefore, model-dependent analyses are to certain
extent unavoidable.

Parity-violating electron scattering has drawn considerable attention
over the recent past as a possible model-independent probe of neutron
distributions. The PREX collaboration
\cite{prex1,prex2,prexprl12} aims to determine the neutron radius in
$^{208}$Pb with 1\% accuracy by measuring the parity-violating asymmetry $A_{\rm
pv}$ with 3\% accuracy in polarized elastic electron scattering performed on a
fixed $^{208}$Pb target. See the contribution to this volume by Horowitz, Kumar,
and Michaels for a detailed account of PREX (and CREX).
Motivated by this landmark experiment, prior to publication of the results of
the first run \cite{prexprl12}, we reported a theoretical study about the
relation of $A_{\rm pv}$ with the NST of $^{208}$Pb and with
the density dependence of the symmetry energy \cite{Roc11}.
Here, we reviewed our study of \cite{Roc11} and extended it in some aspect.
We have computed $A_{\rm pv}$ in DWBA calculations in a comprehensive
set of mean-field models that reproduce the
experimental charge radius of $^{208}$Pb within 1\% accuracy. The
electric and weak charge density distributions have been obtained from
the self-consistent point-like neutron and proton densities of the
nuclear models folded with electromagnetic and weak charge form
factors. An almost perfect linear correlation ($r>0.99$)
between $A_{\rm pv}$ and $\Delta r_{np}$ of $^{208}$Pb is found in the nuclear
mean-field models. Also a very good correlation, but a little less
linear ($r\sim0.97$), exists between $A_{\rm pv}$ and the neutron
radius. The calculations also reveal that a 3\% accuracy in $A_{\rm
pv}$ leads to 1\% accuracy in the neutron radius, hence supporting the goal of
PREX.

A 3\% accuracy in $A_{\rm pv}$ was unachieved in the first run of PREX
but will be within reach of PREX-II \cite{prexprl12,prexii}. Here, as in
Ref.~\cite{Roc11}, we discussed a 3\%-accuracy test value placed at $A_{\rm
pv}=0.715$ ppm. It implies
a central value for $\Delta r_{np}$ of $^{208}$Pb within the estimates obtained
both from a variety of strong probes and from information from observed masses
and radii of neutron stars. Although our central value of $A_{\rm pv}$ is not
the value measured in the first run of PREX ($A_{\rm pv}=0.656\pm(0.060)_{\rm
stat} \pm (0.014)_{\rm syst}$ ppm), they agree within uncertainties.
Our predictions about the accuracy for extracting the neutron radius, neutron
skin, and slope of the symmetry energy if $A_{\rm pv}$ is measured
to 3\% accuracy, remain valid almost independently of the central value of
$A_{\rm pv}$ in the analyzed ranges.

We have found a strong correlation between $A_{\rm pv}$ and the
slope of the symmetry energy $L$. A relatively large spread of
$\pm 40$ MeV in $L$ is predicted if the accuracy of an $A_{\rm pv}$ measurement
is at the 3\% level. Our study supports that an eventual measurement of $A_{\rm
pv}$ with 1\% accuracy would allow to constrain $\Delta r_{np}$ of
$^{208}$Pb and $L$ within narrow bounds close to $\pm 0.02$ fm and $\pm 10$ MeV,
respectively.

Finally, we discussed some features of the size and shape of the
neutron density distribution of $^{208}$Pb according to the mean-field models.
We concluded by discussing calculations of $A_{\rm pv}$ in $^{208}$Pb performed
with the experimental charge density and neutron densities parametrized by 2pF
shapes, where the neutron surface diffuseness was varied within the range
predicted by the theory.

Recently, a new measurement of the electric dipole response in $^{208}$Pb
has allowed to determine the electric dipole polarizability $\alpha_D$ in this
nucleus to 3\% accuracy ($\alpha_D=20.1\pm0.6$ fm$^3$) \cite{Tam11}; see the
contribution to this volume by Tamii, von Neumann-Cosel, and Poltoratska. A skin
$\Delta r_{np} = 0.156^{+0.025}_{-0.021}$ fm in ${}^{208}$Pb was deduced
\cite{Tam11} via the covariance analysis of Ref.~\cite{Rei10} within a Skyrme
model. In a later study using a wide class of functionals, $\Delta
r_{np} = 0.168\pm 0.022$ fm was reported from the $\alpha_D$ data \cite{Pie12}.
A new study of the $\alpha_D$ observable \cite{Roca13a} has been inspired by
insights from the DM. That is, the DM expression for the dipole
polarizability~\cite{meyer82}:
\begin{equation} 
\alpha_D^{\rm DM} = \frac{\pi e^2}{54} \frac{A \langle
r^2\rangle}{J} \left(1+\frac{5}{3}\,\frac{9J}{4Q}A^{-1/3}\right)
\label{dpdm1} 
\end{equation} 
considered together with the DM formula for the neutron skin thickness (cf.\
Eqs.\ (\ref{eq9})--(\ref{eq10}) in Section \ref{sec:2}),
suggests that the dipole polarizability times the symmetry energy
at saturation ($\alpha_D J$) is strongly correlated with $\Delta r_{np}$ in
$^{208}$Pb and with the slope of the symmetry energy $L$
\cite{Roca13a}. These correlations have been confirmed to be
very accurate by self-consistent calculations in a large set of
nuclear functionals \cite{Roca13a}. A similarly strong correlation has been
shown between $\alpha_D J$ and the parity-violating asymmetry $A_{\rm pv}$ in
$^{208}$Pb \cite{Roca13a}. Hence, precise measurements of the dipole
polarizability and the parity-violating asymmetry in $^{208}$Pb, and
in other neutron-rich heavy- and medium-mass nuclei, are expected to offer
trailblazing opportunities both to constrain a fundamental property of nucleon
densities such as the neutron skin thickness and to scrutinize the density
dependence of the symmetry energy around nuclear saturation.

\section*{Acknowledgments}
Our research has been partially funded by the Spanish Consolider-Ingenio 2010 Programme CPAN CSD2007-00042 and by Grants No.\ FIS2011-24154 from MICINN (Spain) and FEDER, No.\ 2009SGR-1289 from Generalitat de Cata\-lu\-nya (Spain), and No.\ DEC-2011/01/B/ST2/03667 from National Science Centre (Poland).

\bibliographystyle{apsrev}
\bibliography{bibliography}

\end{document}